\documentclass[12pt, peerreview , onecolumn]{IEEEtran}

\usepackage{tikz}
\usepackage[sumlimits,intlimits]{amsmath}
\usepackage{amsmath,amsfonts,amssymb}%
\usepackage{bm}%
\usepackage{graphicx,graphics}%

\usepackage{cases}%
\usepackage[noadjust]{cite}%
\usepackage{color}%
\usepackage{cite,url}
\usepackage{verbatim}
\usepackage{algorithm}
\usepackage{balance}
\usepackage{multirow}
\usepackage{multicol}
\usepackage{stfloats}
\usepackage{subfigure}
\usepackage{xspace}
\usepackage{enumerate}
\usepackage{rotating}
\usepackage{bbm}

\newtheorem{remark}{Remark}
\newtheorem{theorem}{Theorem}

\newtheorem{lemma}{Lemma}

\ifCLASSOPTIONpeerreview
\renewcommand{\baselinestretch}{1.65}
\fi

\newcommand{\lh}{\lambda_h}
\newcommand{\lp}{\lambda_p}

\newcommand{\lgg}{\lambda_g}

\newcommand{\sigr}{\sigma^2_{n_r}}
\newcommand{\sigd}{\sigma^2_{n_d}}

\newcommand{\por}{I_{o,r,i}}
\newcommand{\pode}{I_{o,d,i}}
\newcommand{\h}{|h_i|^2}
\newcommand{\g}{|g_i|^2}
\newcommand{\ba}{\bar{a}}
\newcommand{\bb}{\bar{b}}
\newcommand{\bz}{\bar{z}}

\begin{document}

\title{\vspace*{-0pt}{Wireless-Powered Relays in Cooperative Communications: Time-Switching Relaying Protocols and Throughput Analysis}}

\author{\IEEEauthorblockN{Ali A.~Nasir, \textit{Member, IEEE}, Xiangyun Zhou, \textit{Member, IEEE}}, \\ Salman Durrani, \textit{Senior Member, IEEE}, and Rodney A. Kennedy, \textit{Fellow, IEEE}}

\maketitle

{\let\thefootnote\relax\footnotetext{ Ali A. Nasir, Xiangyun Zhou, Salman Durrani, and Rodney Kennedy are with Research School of Engineering, The Australian National University, Canberra, Australia.  Emails: ali.nasir@anu.edu.au, xiangyun.zhou@anu.edu.au, salman.durrani@anu.edu.au, and rodney.kennedy@anu.edu.au. A preliminary version of this work is accepted for presentation at 2015 IEEE ICC in London, UK \cite{Nasir-15-P}. The work of A. A. Nasir, X. Zhou, and S. Durrani was supported in part by the Australian Research Council's Discovery Project funding scheme (project number DP140101133). \vspace{-00pt}}}

\vspace{-0pt}

\vspace{-12pt}

\begin{abstract}
We consider wireless-powered amplify-and-forward and decode-and-forward relaying in cooperative communications, where an energy constrained relay node first harvests energy through the received radio-frequency signal from the source and then uses the harvested energy to forward the source information to the destination node. We propose time-switching based energy harvesting (EH) and information transmission (IT) protocols with two modes of EH at the relay. For continuous time EH, the EH time can be any percentage of the total transmission block time. For discrete time EH, the whole transmission block is either used for EH or IT. The proposed protocols are attractive because they do not require channel state information at the transmitter side and enable relay transmission with preset fixed transmission power. We derive analytical expressions of the achievable throughput for the proposed protocols. The derived expressions are verified by comparison with simulations and allow the system performance to be determined as a function of the system parameters. Finally, we show that the proposed protocols outperform the existing fixed time duration EH protocols in the literature, since they intelligently track the level of the harvested energy to switch between EH and IT in an online fashion, allowing efficient use of resources.

\end{abstract}

\begin{keywords}
Wireless communications, amplify-and-forward, decode-and-forward, throughput, wireless energy harvesting.
\end{keywords}

\ifCLASSOPTIONpeerreview
    \newpage
\fi

\section{Introduction}

Radio frequency (RF) or wireless energy harvesting has recently emerged as an attractive solution to power nodes in future wireless networks~\cite{Lu-14,Gunduz-14,Huang-15-A,Visser-13-A,Tabassum-15}.
Wireless energy harvesting techniques are now evolving from theoretical concepts into practical devices for low-power electronic applications~\cite{powercast}. The feasibility of wireless energy harvesting for low-power cellular applications has been studied using experimental results, which have been summarised in~\cite{Lu-14}. With wireless energy harvesting, there is a choice between harvesting energy from ambient sources or by carefully designing wireless power transfer links. For instance, a power density of around 1 mW/m$^2$ is reported around $50$ meter distance from the base station in the GSM band (935 MHz - 960 MHz)~\cite{Huang-15-A}, which means that a wireless device with a typical size of around 100 cm$^2$ can harvest power in the range of tens of $\mu$W. Such an amount of harvested power could be sufficient for relaying nodes in sensor networks with sporadic activities. For devices that need to support frequent communication activities, harvesting ambient RF energy is not sufficient. Instead, harvesting wireless energy from carefully designed power transfer links is needed. In addition, the energy conversion efficiency of the wireless energy harvesting plays an important role in determining the amount of energy that can be harvested. Employing different circuit design technologies, wireless energy harvesting with energy conversion efficiency in the range of $10\%$-$80\%$ has been reported over a wide range of frequencies, e.g., 15 MHz - 2.5 GHz \cite{Lu-14,Shinohara-14-B}. More specifically, energy conversion efficiency of around $65\%$ has been reported in the ISM band (900 MHz, 2.4 GHz) with 13 nm CMOS technology \cite{Lu-14}.

A unique advantage of RF energy harvesting lies in the fact that RF signals can carry both energy and information, which enables energy constrained nodes to scavenge energy and receive information~\cite{Varshney-08-P,Grover-10-P}. The studies in \cite{Varshney-08-P,Grover-10-P} implicitly assumed that the received energy can still be harvested after passing through an information decoder, which is practically infeasible given the limitations of the current state-of-the-art of electronic circuits \cite{Huang-13-A}. This motivated the design of practically realizable receiver designs which separate information decoding and energy harvesting processes using time-switching or power splitting~\cite{Zhou-12-A}. Such designs have been widely adopted in the recent literature~\cite{Xiang-12-A,Zhang-12-A,Liu-12-A,Nguyen-15-P}.

An important application of wireless energy harvesting is in cooperative relaying, where an intermediate relay node assists in the transmission of the source information to the destination. The relay node may have limited battery reserves and thus relies on some external charging mechanism in order to remain active in the network \cite{Medepally-10-A}. Therefore, energy harvesting in such networks is particularly important as it can enable information relaying. Recent survey papers have discussed many application scenarios, e.g., emerging ultra-dense small scale cell deployments, sensor networks and extremely dense wireless networks, where a combination of energy harvesting and relaying can be useful and practical~\cite{Lu-14,Gunduz-14,Huang-15-A,Visser-13-A,Tabassum-15}.

\subsection{Related Work}

The majority of the research in wireless energy harvesting and information processing has considered point-to-point communication systems and studied rate-energy trade-off assuming single-input-single-output (SISO) \cite{Grover-10-P,Zhou-12-A,Liu-12-A,Popovski-12-A,Xu-12-A}, single-input-multiple-output (SIMO) \cite{Liu-13-A-DPST}, multiple-input-single-output (MISO) \cite{Shen-13-A}, and multiple-input-multiple-output (MIMO) \cite{Zhang-12-A,Park-13-A} setups. The application of wireless energy harvesting to orthogonal frequency division multiplexing (OFDM) \cite{Ng-13-P-OFDM} and cognitive radio \cite{Lee-13-A,Nguyen-15-P} based systems has also been considered. Energy beamforming through wireless energy harvesting was studied for the multi-antenna wireless broadcasting system in \cite{Xiang-12-A,Shi-13-A}. Secure transmission in the presence of eavesdropper under wireless energy harvesting constraint was studied in MISO beamforming systems \cite{Ng-13-P-MISO}. Moreover, multiuser scheduling in the presence of wireless energy harvesting was considered in \cite{Ju-13-A}.

Some studies have recently considered energy harvesting through RF signals in wireless relaying networks \cite{Chalise-12-P,Ishibashi-P-12,Fouladgar-12-A,Krikidis-12-A,Zeng-14-A,Nasir-13-A,Chen-14-P,Krikidis-14-A,Ding-13-A}. The different rate-energy trade-offs to achieve the optimal source and relay precoding in a MIMO relay system was studied in \cite{Chalise-12-P}. The outage performance of a typical cooperative communication system was studied in \cite{Ishibashi-P-12}. However, the authors in \cite{Chalise-12-P,Ishibashi-P-12} assumed that the relay has sufficient energy of its own and does not need external charging. Multi-user and multi-hop systems for simultaneous information and power transfer were investigated in \cite{Fouladgar-12-A}. The optimization strategy in \cite{Fouladgar-12-A} assumed that the relay node is able to decode the information and extract power simultaneously, which, as explained in \cite{Zhou-12-A}, may not hold in practice. Considering amplify-and-forward (AF) relaying under energy harvesting constraints, the outage performance of half-duplex and throughput performance of full-duplex relaying networks was studied in \cite{Krikidis-12-A} and \cite{Zeng-14-A}, respectively. However, perfect channel knowledge for the relay-to-destination link at the relay transmitter was assumed in \cite{Krikidis-12-A} and \cite{Zeng-14-A}. Further, full-duplex relaying as in [30], introduces additional complexity at the relay node due to multiple antenna deployment and the requirement of self interference cancelation.

Recently, considering AF relaying, the throughput performance of a single-way relaying network \cite{Nasir-13-A} and outage probability and ergodic capacity of two-way relaying network \cite{Chen-14-P} under energy harvesting constraints were studied. The outage performance and relay selection criteria in a large scale network with wireless energy harvesting and DF relaying was studied in \cite{Krikidis-14-A}. Finally, for a decode-and-forward (DF) relaying network, the power allocation strategies and outage performance under energy harvesting constraints was studied in \cite{Ding-13-A}. However, \cite{Chen-14-P}, \cite{Krikidis-12-A}, \cite{Krikidis-14-A}, and \cite{Ding-13-A} do not investigate analytical expressions for the achievable throughput at the destination node. In addition, \cite{Nasir-13-A} considers energy harvesting time to have fixed duration and similar to \cite{Chen-14-P}, \cite{Krikidis-14-A}, and \cite{Ding-13-A}, does not allow energy accumulation at the relay node. \textit{Hence, it is still an open problem to design protocols which allow energy accumulation at the relay, i.e., the extra energy, if harvested, is accumulated and stored for future use}.

\subsection{Contribution}

In this paper, considering both AF and DF relaying, we propose protocols for wireless-powered relay nodes in cooperative communications with energy accumulation at the relay. We adopt time-switching (TS) receiver architecture~\cite{Zhou-12-A}, for separate EH and information processing at the relay node. We assume that the energy constrained relay continues to harvest energy from the RF signal transmitted by the source node until it has harvested sufficient energy to be able to transmit the source information with a fixed preset transmission power. Consequently, the duration of EH time at the relay node is variable in the proposed protocols and it depends on the accumulated harvested energy and the quality of the source-to-relay channel. The main contributions of this work are summarized below:
\begin{itemize}

\item Considering energy-constrained AF and DF relaying and block-wise transmission over quasi-static fading channels, we propose TS-based protocols for wireless EH and IT with two different modes of EH. In \textit{continuous time EH}, the duration of EH time can be any percentage of the transmission block time. In \textit{discrete time EH}, the whole transmission block is either used for EH or IT. The proposed protocols are attractive because they (i) do not require channel state information at the transmitter side, (ii) enable relay transmission with preset fixed transmission power and (iii) allow energy accumulation at the relay node.

\item We derive analytical expressions of the achievable throughput for the proposed protocols (cf. Theorems \ref{Th1}-\ref{Th4}). The derived expressions allow the system performance to be accurately determined as a function of the system parameters, such as the relay transmission power and the noise powers.

\item We show that the proposed protocols outperform the existing fixed time duration EH protocols in the literature, since they intelligently track the level of the harvested energy to switch between EH and IT in an online fashion, allowing efficient use of resources.

\end{itemize}

The rest of the paper is organized as follows. Section \ref{sec:sys_mod} presents the system model and assumptions. Sections \ref{sec:AF} and \ref{sec:DF} discuss the proposed protocols and analytically derive the achievable throughput for the energy constrained AF and DF relaying, respectively. Section \ref{sec:sim} presents the numerical results. Finally, Section \ref{sec:conclusions} concludes the paper.

\ifCLASSOPTIONpeerreview
\renewcommand{\baselinestretch}{1.75}
\fi

\section{System Model and Assumptions}\label{sec:sys_mod}

We consider a cooperative communication scenario where a source node, $\mathbb{S}$, communicates with a destination node, $\mathbb{D}$, via an intermediate relay node, $\mathbb{R}$~\cite{Hasna-02-P}. The source and destination nodes are not energy constrained, while the relay node is energy constrained. We assume that there is no direct link between the source and the destination, e.g., due to physical obstacles, which is a valid assumption in many real-world communication scenarios~\cite{Chalise-12-P,Fouladgar-12-A,Krikidis-12-A,Zeng-14-A,Nasir-13-A,Chen-14-P,Ding-13-A}. Thus, deploying EH relay nodes is an energy efficient solution to enable the communication between source and destination when there is no direct link between the source and the destination.\footnote{Note that when a direct link between source and destination is available then there is very little benefit of deploying an energy constrained relay node. This is because the throughput of the direct link  between source and destination is far greater than the throughput of the two-hop link from source to energy constrained relay node to destination.}

\subsubsection{Relay Model} We make the following assumptions regarding the energy constrained relay node:
\begin{itemize}

\item The processing power required by the transmit/receive circuitry at the relay is negligible as compared to the power used for signal transmission from the relay to the destination. This assumption is justifiable in practical systems when the transmission distances are not too short, such that the transmission energy is the dominant source of energy consumption~\cite{Medepally-10-A,Zhou-12-A,Zeng-14-A,Nasir-13-A,Chen-14-P}.

\item The energy constrained relay node first harvests sufficient energy from the source, which is transmitting with power $P_s$, before it can relay the source information to the destination with a preset fixed transmission power $P_r$. For analytical tractability, we assume that the harvested energy at the relay is stored in an infinite capacity battery~\cite{Xu-12-A,Zeng-14-A,Chen-14-P,Ding-13-A}.

\item The relay receiver has two circuits to perform energy harvesting and information processing separately and it adopts the time-switching (TS) receiver architecture~\cite{Nasir-13-A}, i.e., the relay node spends a portion of time for energy harvesting (EH) and the remaining time for information transmission (IT). Such a receiver architecture is widely adopted in the literature~\cite{Zhou-12-A,Xiang-12-A,Zhang-12-A,Liu-12-A,Nguyen-15-P}.

\end{itemize}


\subsubsection{Channel Model} We assume that the $\mathbb{S}$ $\to$ $\mathbb{R}$ and $\mathbb{R}$ $\to$ $\mathbb{D}$ channel links are composed of large scale path loss and statistically independent small-scale Rayleigh fading.\footnote{Note that the proposed protocols are valid for any distribution model assumed for channel fading, however, the closed-form throughput expressions are only valid for Rayleigh distributed channel gains.} We denote the distances between $\mathbb{S}$ $\to$ $\mathbb{R}$ and $\mathbb{R}$ $\to$ $\mathbb{D}$ by $d_1$ and $d_2$, respectively. The $\mathbb{S}$ $\to$ $\mathbb{R}$ and $\mathbb{R}$ $\to$ $\mathbb{D}$ fading channel gains, denoted by $h$ and $g$, respectively, are modeled as quasi-static and frequency non-selective parameters. Consequently, the complex fading channel coefficients $h$ and $g$ are circular symmetric complex Gaussian random variables with zero mean and unit variance. We make the following assumptions regarding the channels:
\begin{itemize}
\item The fading channel gains are assumed to be constant over a \emph{block time} of $T$ seconds and independent and identically distributed from one block to the next. The use of such channels is motivated by prior research in this field~\cite{Zhou-12-A,Zhang-12-A,Medepally-10-A,Liu-12-A,Chalise-12-P,Krikidis-12-A,Ishibashi-P-12,Ding-13-A,Zeng-14-A,Nasir-13-A,Chen-14-P}.
\item The channel state information (CSI) is not available at the transmitter side, while the CSI is obtained at the receiver side through channel estimation at the start of each block. Note that AF relaying does not require the relay to estimate the $\mathbb{S}$ $\to$ $\mathbb{R}$ channel, while DF relaying requires it for decoding purpose.
\item The receiver side CSI is assumed to be perfect, which is in line with the previous work in this research field~\cite{Chalise-12-P,Krikidis-12-A,Ishibashi-P-12,Ding-13-A,Zeng-14-A,Nasir-13-A,Chen-14-P}.
\end{itemize}

\ifCLASSOPTIONpeerreview

\else

\begin{table*}[t]
\vspace{1cm}
\caption{List of important symbols.} \centering
\begin{tabular}{|l|p{6.5in}|} \hline
\multicolumn{1}{|l|} {Symbol} & \multicolumn{1}{l|} {Definition} \\ \hline
$E_o$ & Available harvested energy at the start of any EH-IT pattern. \\
$E_i(0)$ & Available harvested energy at the start of the $i$th block.  \\
$E_i(t)$ & Harvested energy at the time instant $t$ in the $i$th block. \\
$E_i^{0 \to T}$  & The amount of harvested energy per EH block.  \\
$t_o$ & EH-IT pattern start time.\\
$X$ & Random number of successive EH blocks, occurring \textit{before} sufficient amount of energy has been harvested for IT, in discrete time EH (both AF and DF). \\
$Y$ & Random number of successive EH blocks due to relay outage, occurring \textit{after} sufficient amount of energy has been harvested for IT, in discrete time EH (DF only). \\
$n$ & Random number of successive EH blocks due to relay outage, occurring irrespective of accumulated energy, in continuous time EH (DF only). \\ \hline
 \end{tabular}
\label{tab:1}
\end{table*}

\fi

\subsubsection{Protocols} We consider the communication in blocks of $T$ seconds, where in general each block can be composed of two parts, (i) EH and (ii) IT which is split equally into $\mathbb{S}$ $\to$ $\mathbb{R}$ IT and $\mathbb{R}$ $\to$ $\mathbb{D}$ IT. We introduce a parameter, $\alpha_i$, to denote the fraction of the block time allocated for EH in the $i$th block. Thus, the time durations for EH, $\mathbb{S}$ $\to$ $\mathbb{R}$ IT and $\mathbb{R}$ $\to$ $\mathbb{D}$ IT are $\alpha_iT$ seconds, $\frac{(1 - \alpha_i) T}{2}$ seconds and $\frac{(1 - \alpha_i) T}{2}$ seconds, respectively. Due to quasi-static fading, the data packet size is assumed to be very small compared to the block time such that many packets can be transmitted within a block time. Also the decision to switch between EH and IT is carried out in an online fashion which depends on the amount of available harvested energy and the preset value of relay transmit power $P_r$. We define two modes of EH:
\begin{enumerate}
\item continuous time EH in which the relay only harvests that much energy within each block that is needed for its transmission, i.e., $\alpha_i \in (0,1)$.

 \item discrete time EH in which the whole block is used for either EH or IT, i.e., either $\alpha_i=1$ (there is no IT and the whole block is used for EH) or $\alpha_i=0$ (there is no EH and the whole block is used for IT).
\end{enumerate}

Consequently, depending on the mode of EH (continuous or discrete) and the type of relaying protocol (both amplify-and-forward (AF) and decode-and-forward (DF) are considered in this work), different cases can arise for the amount of accumulated harvested energy and the combinations of EH and IT blocks that can occur during the transmission. For a tractable analysis of these cases, we define the following terms and symbols:
\begin{itemize}
\item For \textit{continuous time EH}, an \emph{EH-IT block} is a special block which contains EH for some portion of the block time and IT for the remaining portion of the block time.

\item For \textit{continuous time EH}, an \emph{EH-IT pattern} is a pattern of blocks which either consists of a single EH-IT block or contains a sequence of EH blocks followed by an EH-IT block.

\item For \textit{discrete time EH}, an \emph{EH-IT pattern} is a pattern of blocks which either consists of a single IT block or contains a sequence of EH blocks followed by an IT block.

\item The harvested energy at the time instant $t$ in the $i$th block is denoted as $E_i(t)$. This variable is used to track the harvested energy within a block.

\item The available harvested energy at the start of the $i$th block is denoted as $E_i(0)$. The available harvested energy at the start of any EH-IT pattern is denoted as $E_o$. Note that $E_o$ has a value only at start of any EH-IT pattern, while $E_i(0)$ has a value at the start of every block.

\end{itemize}

In the following sections, we propose TS-based EH and IT protocols with continuous time and discrete time EH in AF and DF relaying. We use the \emph{throughput efficiency} as the figure of merit. It is defined as the fraction of the block time used for successful information transmission, on average, where successful transmission implies the successful decoding of the source message at the destination. Note that we use the terms throughput efficiency and throughput interchangeably in the paper. We consider AF and DF relaying separately because the throughput analysis and proposed protocols are different for both systems.

For the sake of convenience, we use the same symbolic notation for different symbols, e.g., throughput $\tau$, energy harvesting time, $\alpha$, outage probability, $p_{o}$, etc. in Sections \ref{sec:AF} and \ref{sec:DF} for AF and DF relaying, respectively. Hence, the expression for any particular variable can be found in the relevant section. For notational convenience, the commonly used symbols in Sections \ref{sec:AF} and \ref{sec:DF} are summarized in Table \ref{tab:1}.
%


\ifCLASSOPTIONpeerreview

\begin{table}[t]
\caption{List of important symbols.} \centering
\begin{tabular}{|l|p{6.5in}|} \hline
\multicolumn{1}{|l|} {Symbol} & \multicolumn{1}{l|} {Definition} \\ \hline
$E_o$ & Available harvested energy at the start of any EH-IT pattern. \\
$E_i(0)$ & Available harvested energy at the start of the $i$th block.  \\
$E_i(t)$ & Harvested energy at the time instant $t$ in the $i$th block. \\
$E_i^{0 \to T}$  & The amount of harvested energy per EH block.  \\
$t_o$ & EH-IT pattern start time.\\
$X$ & Random number of successive EH blocks, occurring \textit{before} sufficient amount of energy has been harvested for IT, in discrete time EH (both AF and DF). \\
$Y$ & Random number of successive EH blocks due to relay outage, occurring \textit{after} sufficient amount of energy has been harvested for IT, in discrete time EH (DF only). \\
$n$ & Random number of successive EH blocks due to relay outage, occurring irrespective of accumulated energy, in continuous time EH (DF only). \\ \hline
 \end{tabular}
\label{tab:1}
\end{table}

\else

\fi

\section{Amplify-and-Forward Relaying} \label{sec:AF}

In this section, we first present the mathematical signal model for wireless-powered AF relaying. Then, we propose two protocols and analytically characterize their throughput performance.

\subsection{Signal Model}
The received signal at the relay node, $y_{r,i}$ is given by
\begin{align}\label{eq:yrt}
      y_{r,i} =  \frac{1}{\sqrt{d_1^m}} \sqrt{P_s} h_i s_i  +  n_{r,i},
\end{align}
where $i$ is the block index, $h_i$ is the $\mathbb{S}$ $\to$ $\mathbb{R}$ fading channel gain, $d_1$ is the source to relay distance, $P_s$ is the source transmission power, $m$ is the path loss exponent, $n_{r,i}$ is the additive white Gaussian noise (AWGN) at the relay node, $s_i$ is the normalized information signal from the source, i.e., $\mathbb{E} \{ |s_i|^2 \} = 1$ (which can be realized by any phase shift keying modulation scheme), where $\mathbb{E} \{ \cdot \}$ is the expectation operator and $|\cdot|$ is the absolute value operator.

The AF relay first harvests sufficient energy to be able to transmit with preset power $P_r$ during IT time. Thus, after EH, the AF relay amplifies the received signal and forwards it to the destination. The signal transmitted by the relay, $x_{r,i}$ is given by
\begin{align}\label{eq:xt_AF}
      x_{r,i} =  \frac{\sqrt{P_r} y_{r,i} } { \sqrt{ \frac{P_s |h_i|^2}{d_1^m} + \sigr } },
\end{align}
where $\sigr$ is the variance of the AWGN at the relay node and the factor in the denominator, $\sqrt{\frac{P_s |h_i|^2}{d_1^m} + \sigr}$ is the power constraint factor at the relay. Note that an AF relay can obtain the power constraint factor, $\sqrt{ \frac{P_s |h_i|^2}{d_1^m} + \sigr}$, from the power of the received signal, $y_{r,i}$, and does not necessarily require source to relay channel estimation.

The received signal at the destination, $y_{d,i}$ is given by
\ifCLASSOPTIONpeerreview
\begin{subequations}\label{eq:ydt_AF}
\begin{align}
      y_{d,i} &= \frac{1}{\sqrt{d_2^m}} g_i x_{r,i}  +  n_{d,i} \label{eq:ydt_AF1} \\ &= \underbrace{\frac{\sqrt{P_r P_s} h_i g_i s_i }{ \sqrt{d_2^m} \sqrt{P_s |h_i|^2 + d_1^m \sigr}}}_\text{signal part} + \underbrace{\frac{\sqrt{P_r d_1^m} g_i n_{r,i}}{\sqrt{d_2^m} \sqrt{P_s |h_i|^2 + d_1^m \sigr} } + n_{d,i}}_\text{overall noise}, \label{eq:ydt_AF2}
\end{align}
\end{subequations}
\else
\begin{subequations}\label{eq:ydt_AF}
\begin{align}
      y_{d,i} &= \frac{1}{\sqrt{d_2^m}} g_i x_{r,i}  +  n_{d,i} \label{eq:ydt_AF1} \\ &= \underbrace{\frac{\sqrt{P_r P_s} h_i g_i s_i }{ \sqrt{d_2^m} \sqrt{P_s |h_i|^2 + d_1^m \sigr}}}_\text{signal part}  \notag \\ & \hspace{1.6cm}  + \underbrace{\frac{\sqrt{P_r d_1^m} g_i n_{r,i}}{\sqrt{d_2^m} \sqrt{P_s |h_i|^2 + d_1^m \sigr} } + n_{d,i}}_\text{overall noise}, \label{eq:ydt_AF2}
\end{align}
\end{subequations}
\fi
where \eqref{eq:ydt_AF2} follows from \eqref{eq:ydt_AF1} by substituting $x_{r,i}$ from \eqref{eq:xt_AF} into \eqref{eq:ydt_AF1}, $g_i$ is the $\mathbb{R}$ $\to$ $\mathbb{D}$ fading channel gain, $d_2$ is $\mathbb{R}$ $\to$ $\mathbb{D}$ distance, and $n_{d,i}$ is the AWGN at the destination node.

Using \eqref{eq:ydt_AF}, the signal-to-noise-ratio (SNR) at the $\mathbb{D}$, $\gamma_{d,i}  =  \frac{ \mathbb{E}_{s_i} \{ |\text{signal part in\eqref{eq:ydt_AF2}}|^2\}}{\mathbb{E}_{n_{r,i},n_{d,i}} \{ |\text{overall noise in \eqref{eq:ydt_AF2}}|^2 \}} $, is given by
\begin{align}\label{eq:gD_AF}
      \gamma_{d,i} &=   \frac{ \frac{ P_s P_r |h_i|^2 |g_i|^2 } { d_2^m ( P_s |h_i|^2  + d_1^m \sigr ) } } {   \frac{  P_r |g_i|^2 d_1^m \sigr } { d_2^m ( P_s |h_i|^2  + d_1^m \sigr ) } + \sigd } \notag \\
      & = \frac{P_s P_r |h_i|^2 |g_i|^2 }{P_r |g_i|^2 d_1^m \sigr + d_2^m \sigd ( P_s |h_i|^2  + d_1^m \sigr ) },
\end{align}
where $\sigd$ is the AWGN variance at the destination node.

The $i$th block will suffer from outage if SNR, $\gamma_{d,i}$, is less than the threshold SNR, $\gamma_o$. Thus, the outage indicator, $I_{o,i}$ is given by
\begin{align}\label{eq:pout_AF}
      I_{o,i} = \mathbbm{1}( \gamma_{d,i} < \gamma_o),
\end{align}
where $\mathbbm{1}(\cdot)$ is an indicator function which is equal to $1$ if its argument is true and $0$ otherwise.

In the following subsections, we propose TS-based protocols for EH and IT using both continuous time and discrete time EH.

\subsection{Protocol 1: TS-based Protocol for EH and IT with Continuous Time EH in AF Relaying} \label{sec:AFc}
\ifCLASSOPTIONpeerreview
\begin{figure}[t]
\centering
     \hspace{-.0cm}\includegraphics[scale=1.3]{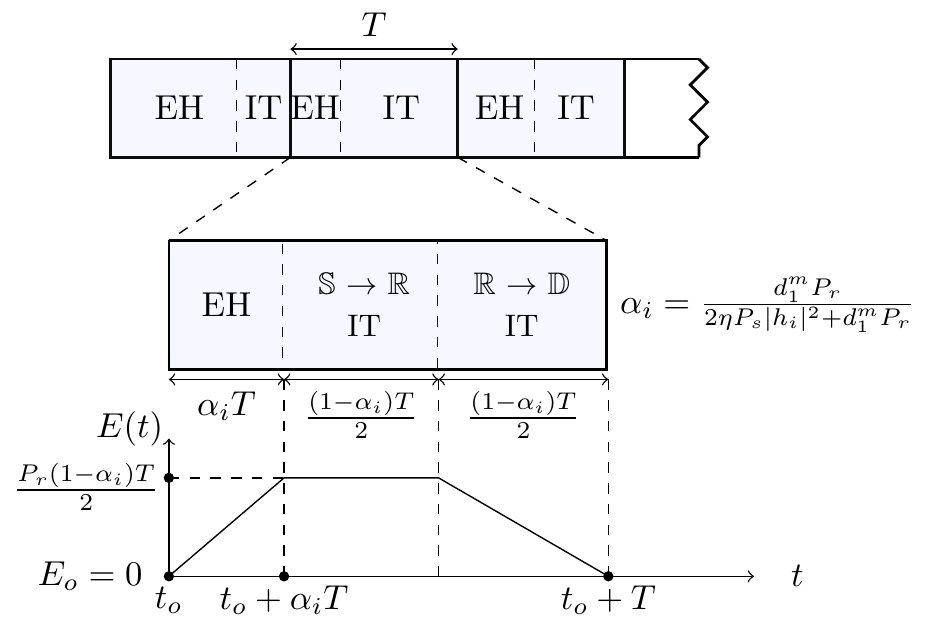}
     \centering
    \caption{Illustration of TS-based protocol for EH and IT in AF relaying with continuous time EH, $\alpha_i \in (0,1)$.}
    \label{fig:AFc}
\end{figure}
\else
\begin{figure}[t]
\centering
     \hspace{-.0cm}\includegraphics[scale=0.97]{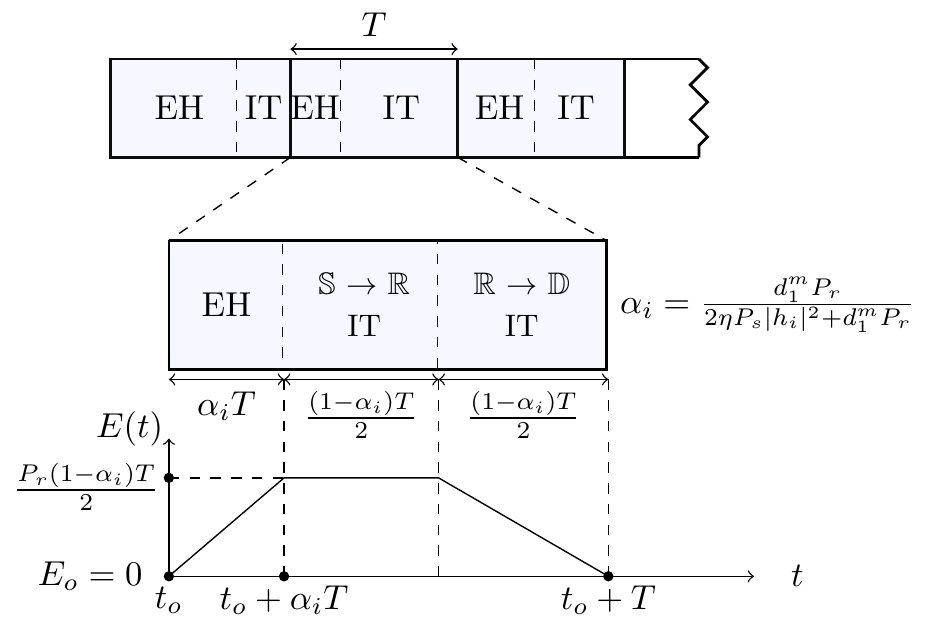}
     \centering
    \caption{Illustration of TS-based protocol for EH and IT in AF relaying with continuous time EH, $\alpha_i \in (0,1)$.}
    \label{fig:AFc}
\end{figure}
\fi

\subsubsection{Description} In this protocol, the relay only harvests that much energy within each block that is needed for its transmission. The continuous time EH in AF relaying guarantees that the relay is able to harvest this required amount of energy within each block (see Remark 1 below \eqref{eq:alpha_t_AFc}). Thus, each EH-IT pattern contains only one block. Within each block, EH occurs for the first $\alpha_i T$ seconds and IT for the remaining $(1-\alpha_i) T$ seconds, where $\alpha_i \in (0,1)$. During IT, half of the time, $\frac{(1-\alpha_i) T}{2}$, is used for $\mathbb{S}-\mathbb{R}$ IT and the remaining half time, $\frac{(1-\alpha_i) T}{2}$, is used for $\mathbb{R}-\mathbb{D}$ IT. Since all of the harvested energy during EH time is consumed during IT time, there is no accumulated harvested energy and $E_o = E_i(0)=0$. The protocol is illustrated in Fig. \ref{fig:AFc}.

\ifCLASSOPTIONpeerreview
\else
\begin{figure*}[t]
\centering
     \hspace{-.0cm}\includegraphics[scale=1.0]{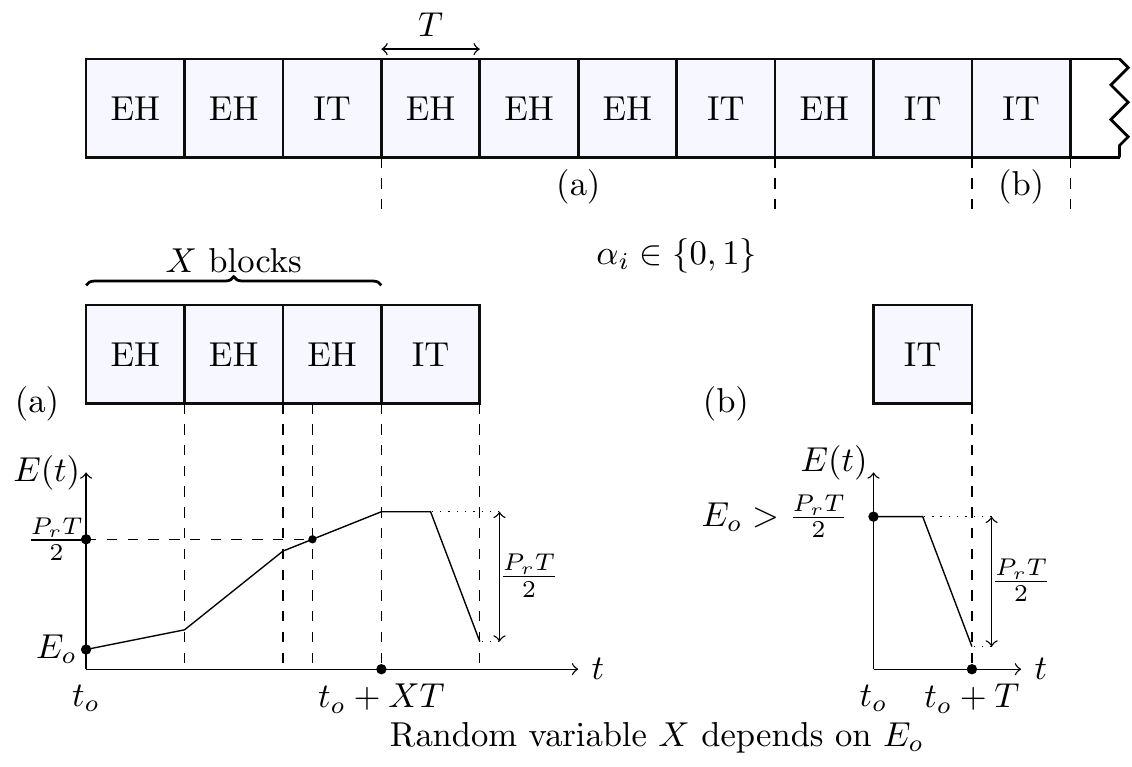}
     \centering
    \caption{Illustration of TS-based protocol for EH and IT in AF relaying network with discrete time EH scheme, $\alpha_i \in \{0,1\}$, for different EH-IT patterns. (a) $X=3$ EH blocks before an IT block, (b) $X=0$ (a single IT block).}
    \label{fig:AFd}
\end{figure*}
\fi

\subsubsection{Energy Analysis} Using \eqref{eq:yrt}, the harvested energy at the time instant, $\alpha_i T$, is given by
\begin{align}\label{eq:EH1_AFc}
      E_i(\alpha_i T) =  \frac{\eta P_s |h_i|^2}{d_1^m} \alpha_i T,
\end{align}
where $0 < \eta < 1$ is the energy conversion efficiency~\cite{Lu-14,Shinohara-14-B}.

The relay needs to forward the source message to the destination with the preset value of the relay power, $P_r$, within the time span, $\frac{(1-\alpha_i) T}{2}$. Thus, the harvested energy at the time instant $\alpha_iT$ is given by
\begin{align}\label{eq:EH2_AFc}
      E_i(\alpha_i T) =  P_r \frac{(1-\alpha_i) T}{2}.
\end{align}
By equating \eqref{eq:EH1_AFc} and \eqref{eq:EH2_AFc}, we can find the value of $\alpha_i$ as
\begin{align}\label{eq:alpha_t_AFc}
     \alpha_i = \frac{d_1^m P_r}{2\eta P_s | h_i|^2  + d_1^m P_r }.
\end{align}

\ifCLASSOPTIONpeerreview
\else
\begin{figure*}[t]
\centering
     \hspace{-.0cm}\includegraphics[scale=1.0]{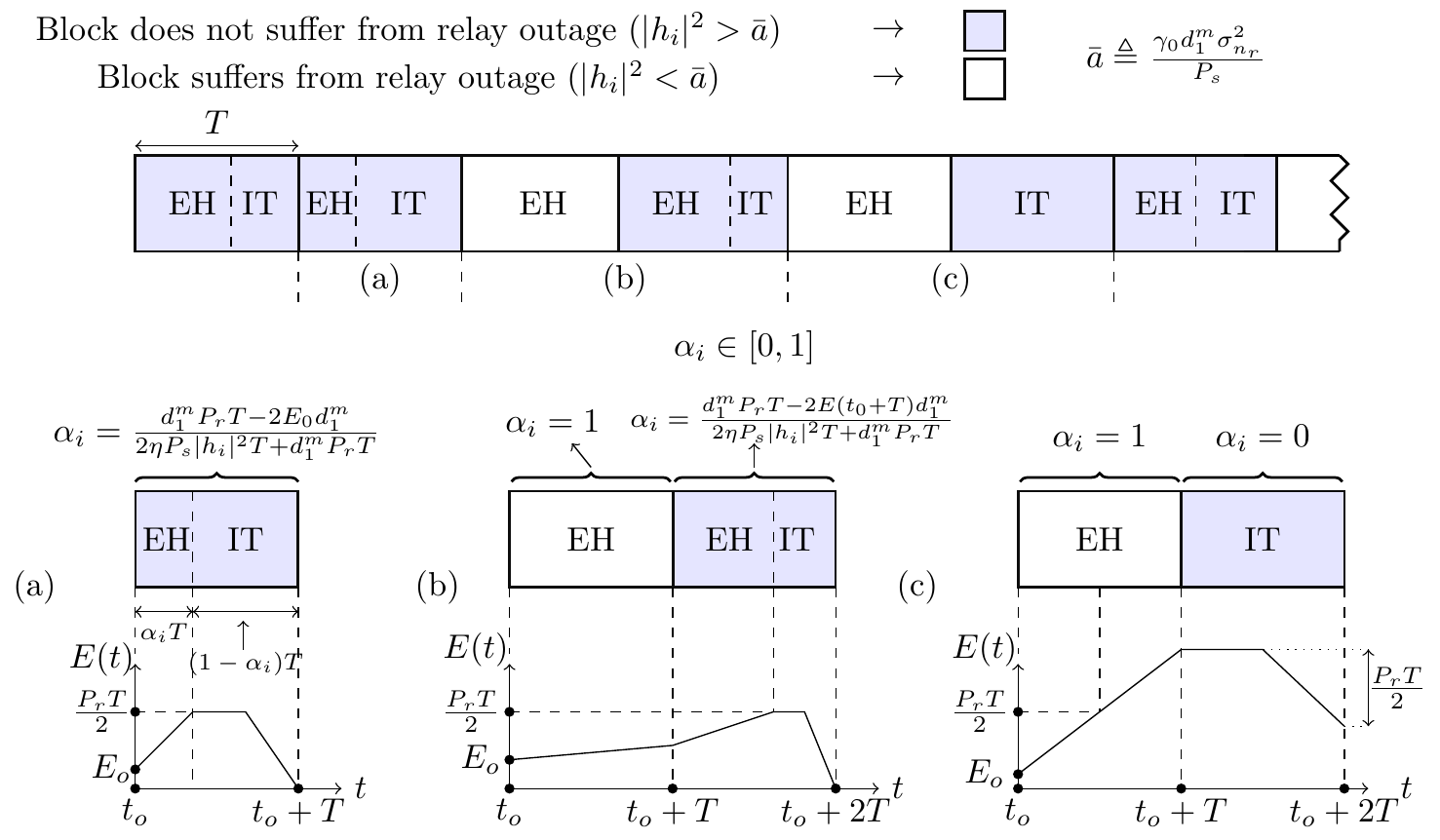}
     \centering
    \caption{Illustration of TS-based protocol for EH and IT in DF relaying, with continuous time EH, $\alpha_i \in [0,1]$ for three different EH-IT patterns (a) single EH-IT block (b) $n=1$ EH block due to relay outage followed by an EH-IT block and (c) $n=1$ EH block due to relay outage followed by an IT block. $n$ is the random number of successive EH blocks due to relay outage, occurring irrespective of accumulated energy, in continuous time EH (DF only).}
    \label{fig:DFc}
\end{figure*}
\fi

\begin{remark}
The relay is always able to harvest the required amount of energy to be able to transmit with preset power $P_r$, within each block. This can be proven as follows. We can observe from \eqref{eq:alpha_t_AFc} that the term $d_1^m P_r$ is common in the numerator and the denominator. This implies that $\alpha_i$ is always between 0 and 1, which means that the relay will always be able to harvest the required amount of energy, that would enable IT by the relay node with preset power $P_r$, within each block. The relay fixed preset power can be appropriately chosen to achieve a desired quality of service (e.g., maximum throughput). The number of packets sent by the relay in each block is variable because it depends on the energy harvesting time $\alpha_i$, which in turn depends on the quality of the source-to-relay channel. If $\alpha_i$ is large, leaving smaller amount of block time for IT, then a smaller number of packets will be transmitted. If $\alpha_i$ is small, leaving larger amount of block time for IT, then a larger number of packets will be transmitted. The throughput definition adopted in our paper corresponds to the portion of the block time that supports reliable IT with no outage. In addition, since a finite number of packets can be sent over the block time, mathematically $\alpha_i$ can only take values from a discrete set. However, due to quasi-static fading channel assumption, the packet size can be very small compared to the block time and thus, $\alpha_i$ can be approximately treated as a continuous variable for continuous time EH protocol.
\end{remark}

%

\subsubsection{Throughput Analysis} Given that $\frac{(1 - \alpha_i) T }{2}$ is the effective communication time within the block of time $T$ seconds, the throughput, $\tau_i$ is given by
\begin{align}\label{eq:tau_AFc}
      \tau_i = (1 - I_{o,i})  \frac{(1 - \alpha_i) T /2 }{T} = \frac{(1 - I_{o,i}) (1 - \alpha_i)}{2},
\end{align}
where the outage indicator, $I_{o,i}$, and energy harvesting time, $\alpha_i$ are defined in \eqref{eq:pout_AF} and \eqref{eq:alpha_t_AFc}, respectively, and $\tau_i$ is a function of the channel gains $h_i$ and $g_i$.

We analytically evaluate the throughput and the main result is summarized in Theorem~\ref{Th1} below.
\begin{theorem}\label{Th1}
The throughput for the TS-based protocol for EH and IT with continuous time EH in AF relaying is given by
\begin{align} \label{eq:tau_AFc_ana}
\tau = \mathbb{E}_{h_i,g_i} \{ \tau_i \} = \frac{e^{-  \frac{a+d}{c} }}{2} \left( u K_{1}(u) -  c d_1^m P_r \nu \right),
\end{align}

\noindent where $a \triangleq P_s d_2^m \sigd \gamma_o$, $c \triangleq P_s P_r $, $ d \triangleq P_r d_1^m \sigr \gamma_o$, $\nu \triangleq \int_{x=0}^{\infty} \frac{e^{-\left( x + \frac{ad+bc}{c^2 x} \right)}}{2 c \eta P_s x + 2 \eta P_s d + c d_1^m P_r} dx$, $b \triangleq d_1^{m} d_2^m \sigr \sigd \gamma_o$, $u \triangleq \sqrt{\frac{4(ad+bc)}{c^2} }$, and $K_{1}(\cdot)$ is the first-order modified Bessel function of the second kind \cite{Gradshteyn-80-B}.
\end{theorem}

\begin{IEEEproof}
See Appendix \ref{app:A}.
\end{IEEEproof}

\begin{remark}\label{R00}
It does not seem tractable to find the closed-form result for $\nu$ because of the constant term $2 \eta P_s d + c d_1^m P_r$ in the denominator. However, for given system parameters, $P_s$, $P_r$, $\sigr$, $\sigd$, $d_1^{m}$, $d_2^m$, $R$, and $\eta$, the value of $\nu$, needed in \eqref{eq:tau_AFc_ana}, can be easily obtained through numerical integration.
\end{remark}

\subsection{Protocol 2: TS-based Protocol for EH and IT with Discrete Time EH in AF Relaying} \label{sec:AFd}
\ifCLASSOPTIONpeerreview
\begin{figure}[t]
\centering
     \hspace{-.0cm}\includegraphics[scale=1.3]{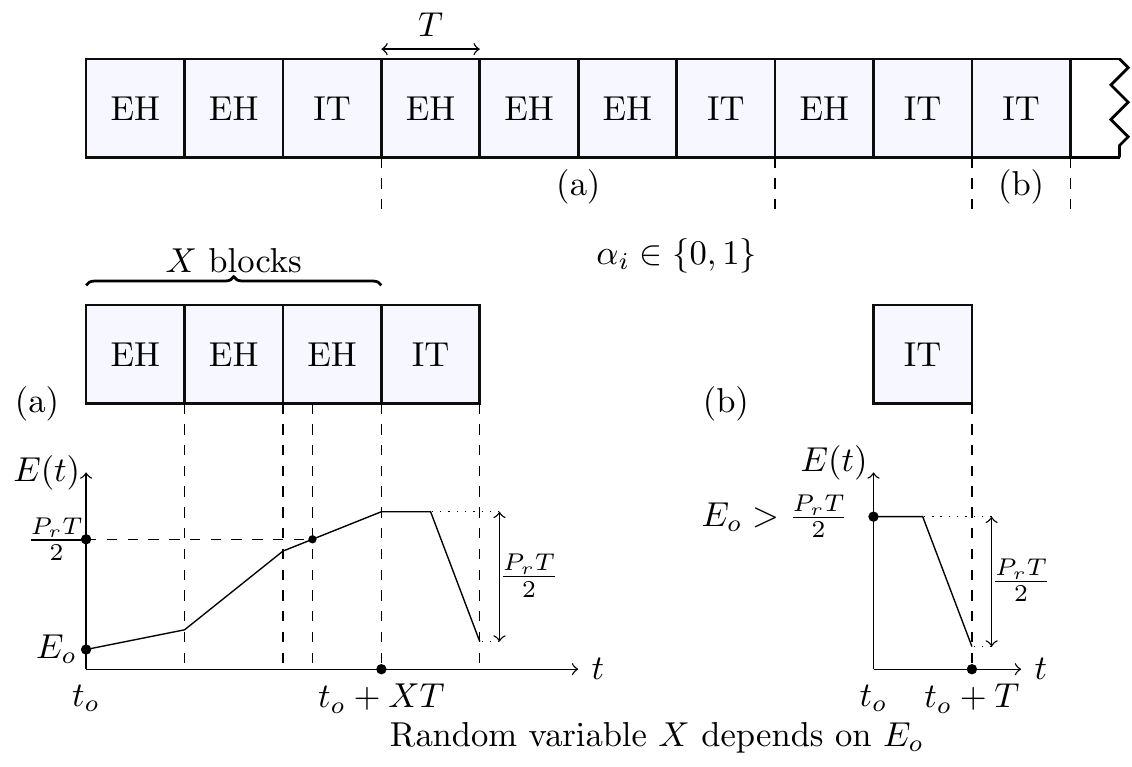}
     \centering
    \caption{Illustration of TS-based protocol for EH and IT in AF relaying network with discrete time EH scheme, $\alpha_i \in \{0,1\}$, for different EH-IT patterns. (a) $X=3$ EH blocks before an IT block, (b) $X=0$ (a single IT block).}
    \label{fig:AFd}
\end{figure}
\else
\fi

\subsubsection{Description}\label{sec:XXX} In this protocol, each block is dedicated either for EH ($\alpha_i = 1$) or IT ($\alpha_i = 0$). If $E_i(0) > \frac{P_r T}{2}$\footnote{This is because the relay needs to transmit the source information for $T/2$ time with the power $P_r$.}, the block is used for IT, otherwise it is used for EH. During IT, half the time, $\frac{T}{2}$, is used for $\mathbb{S}-\mathbb{R}$ IT and the remaining half time, $\frac{T}{2}$, is used for $\mathbb{R}-\mathbb{D}$ IT. Consequently, in this protocol two types of EH-IT patterns are possible:
\begin{itemize}
\item EH-IT pattern contains $X$ successive EH blocks before an IT block, where $X$ is a discrete random variable which depends on $E_o$. Note that an EH-IT pattern having a large $E_o$ is more likely to have a small number of $X$ blocks.
\item EH-IT pattern contains a single IT block because $E_i(0) > \frac{P_r T}{2}$.
\end{itemize}
\noindent These two types of EH-IT patterns are illustrated in Fig.~\ref{fig:AFd}, where in Fig.~\ref{fig:AFd} (a) the desired energy level is shown to be achieved somewhere in the middle of the third EH block and the EH-IT pattern contains $X=3$ EH blocks before an IT block. Note that since the harvested energy during $X$ EH blocks can exceed $\frac{P_r T}{2}$ and IT block only consumes $ \frac{P_r T}{2}$ energy, $E_o$ can have a value greater than $0$ in this protocol.
\begin{remark}
In Protocol 2, the relay only needs to check its available energy at the start of each block. This is in contrast to Protocol 1, where the relay needs to continuously check its available energy during each block. In both protocols, once the harvested energy is sufficient for IT, it is assumed that the relay sends 1 bit to the source and destination nodes to indicate the start of IT. In this work, for the sake of tractability, we assume that this control channel information is error free.
\end{remark}

\subsubsection{Energy Analysis} If a block is used for EH, using \eqref{eq:yrt}, the total harvested energy during the block, denoted by $E_i^{0 \to T}$, is given by
\begin{align}\label{eq:EH1_AFd}
      E_i^{0 \to T} =  \frac{\eta P_s |h_i|^2}{d_1^m} T.
\end{align}

The value of $\alpha_i$ and $E_i(T)$ for the $i$th block is given by
\begin{align}\label{eq:alpha_t_AFd}
      \alpha_i = \begin{cases}
                    1 , & E_i(0) < \frac{P_r T}{2} \\
                    0 , & E_i(0) > \frac{P_r T}{2}
       \end{cases}
\end{align}
\begin{align}\label{eq:Ei_AFd}
       E_i(T) = \begin{cases}
                    E_i(0) + E_i^{0 \to T} = E_i(0) + \frac{\eta P_s |h_i|^2}{d_1^m} T , & \alpha_i=1 \\
                    E_i(0) - \frac{P_r T}{2} , & \alpha_i=0
       \end{cases}
\end{align}

\subsubsection{Throughput Analysis}\label{sec:AFd_ana}
Using $\alpha_i$ from \eqref{eq:alpha_t_AFd} and $I_{o,i}$ from \eqref{eq:pout_AF}, the general block throughput is given by $\tau_i = \frac{(1 - I_{o,i}) (1 - \alpha_i)}{2} $, which in addition to $h_i$ and $g_i$, is also a function of $\{h_{i-1},h_{i-2},\hdots\}$ because of the variable $E_i(0)$. In order to determine the throughput, we need to determine the distribution of $E_o$ and the conditional distribution of $\bar{X} \triangleq X-1$, given the value of $E_o$, where $\bar{X}$ denotes the number of EH blocks that arrive within the energy interval $\left(E_o,\frac{P_r T}{2}\right]$. These are given in the lemmas below.

\begin{lemma}\label{L1}
The available harvested energy available at the start of any EH-IT pattern, $E_o$, is exponentially distributed with mean $\rho$ and the probability density function (PDF)
\begin{align}\label{eq:f_Eo_AFd}
f_{E_o}(\epsilon) =  \frac{1}{\rho } e^{- \frac{\epsilon}{\rho } }, \;\; \epsilon > 0
\end{align}
where $\rho \triangleq \frac{\eta P_s T}{d_1^m}$.
\end{lemma}
\begin{IEEEproof}
See Appendix \ref{app:AB}.
\end{IEEEproof}
\begin{lemma}\label{L2}
Given the value of $E_o$, if $E_o < \frac{P_r T}{2}$, $\bar{X} \triangleq X-1$ is a Poisson random variable with parameter $ \lp$ and the probability mass function (PMF) of $\bar{X}$ is given by the Poisson PMF,
\begin{align}\label{eq:pX_AFd}
p_{\bar{X}|E_o}(\bar{x}|E_o) = \frac{\lp^{\bar{x}} \hspace{0.2cm}  e^{-\lp}}{\bar{x}!}, \;\;\;\; E_o < \frac{P_r T}{2}
\end{align}
where $  \lp \triangleq \mathbb{E}\left\{ \bar{X}|E_o \right\} = \frac{1}{\rho} \left( \frac{P_r T}{2} - E_o \right)$ and $\bar{X} \in \{0,1,2,\hdots\}$. If $E_o \ge \frac{P_r T}{2}$, $X=0$, i.e., it is a constant.
\end{lemma}
\begin{IEEEproof}
$\bar{X} = X-1$ denotes the number of EH blocks that arrive within the energy interval $\left(E_o,\frac{P_r T}{2}\right]$. Since the energy harvested per EH block, $E_i^{0 \to T}$, is exponentially distributed with mean $\rho$ (see Appendix B), the number of EH blocks required to harvest the required energy, $\frac{P_r T}{2}-E_o$, follows a Poisson random variable with parameter $\frac{1}{\rho} \left( \frac{P_r T}{2} - E_o \right)$ \cite[Ch. 2]{Gallager-96-B}. Consequently, $\bar{X}$, given $E_o$, is a Poisson random variable with the Poisson PMF defined in \eqref{eq:pX_AFd} \cite[Theorem 2.2.4]{Gallager-96-B}.
\end{IEEEproof}

Using Lemmas~\ref{L1} and~\ref{L2}, we can derive the throughput result for Protocol 2, which is given in Theorem~\ref{Th2} below.

\begin{theorem}\label{Th2}
The throughput for the TS-based Protocol for EH and IT with discrete time EH in AF relaying is given by
\begin{align}\label{eq:tau_AFd_ana}
\tau = \mathbb{E}_{g_i,\mathbf{h}} \{ \tau_i \} = \frac{e^{- \frac{a+d}{c} } u K_{1} \left(  u \right)}{2 \left( 1+\frac{P_r d_1^m}{2 \eta P_s } \right)},
\end{align}
where $\mathbf{h} \triangleq \{h_i,h_{i-1},h_{i-2},\hdots\}$, $\tau_i$ is given below \eqref{eq:Ei_AFd}, $a$, $b$, $c$, $d$, and $u$ are defined below \eqref{eq:tau_AFc_ana}, and $K_{1}(\cdot)$ is defined in Theorem \ref{Th1}.
\end{theorem}
\begin{IEEEproof}
See Appendix \ref{app:B}.
\end{IEEEproof}

\section{Decode-and-Forward Relaying} \label{sec:DF}

In this section, we first present the mathematical signal model for wireless-powered DF relaying. Then, we propose two protocols and analytically characterise their throughput performance. \emph{In contrast to AF relaying, IT in DF relaying is affected by whether the source to relay channel is in outage. If the source to relay channel is in outage (\textit{for simplicity, we refer to this as relay outage}), the relay will alert the source and destination nodes by sending a single bit, and EH will be carried out for that entire block.}

\subsection{Signal Model}

The received signal at $\mathbb{R}$, $y_r(t)$, is given by \eqref{eq:yrt}. During IT, the received signal at $\mathbb{D}$, $y_{d,i}$, is given by
\begin{align}\label{eq:ydt_DF}
      y_{d,i} =  \frac{1}{\sqrt{d_2^m}} \sqrt{P_r} g_i s_i  +  n_{d,i},
\end{align}
Using \eqref{eq:yrt} and \eqref{eq:ydt_DF}, the SNR at the relay and destination nodes are given by
\begin{subequations}\label{eq:gR_gD}
\begin{align}
      \gamma_{r,i} &=  \frac{P_s \h}{d_1^m \sigr} \\
      \gamma_{d,i} &=  \frac{P_r \g}{d_2^m \sigd} .
\end{align}
\end{subequations}
Using \eqref{eq:gR_gD}, the outage indicator is given by
\begin{align}\label{eq:pout_DF}
     I_{o,i} =  \mathbbm{1}( \gamma_{d,i} < \gamma_o ) = \mathbbm{1}(\g < \bb)
\end{align}

\ifCLASSOPTIONpeerreview

\else
 \newcounter{MYtempeqncnt}
 \begin{figure*}[!b]
  \hrulefill
 \vspace{-0.0cm}
\normalsize
\setcounter{MYtempeqncnt}{19}
\setcounter{equation}{24}
\begin{align}\label{eq:tau_DFc_ana}
\tau  &\ge   \frac{e^{-(\ba+\bb)}}{4 \eta P_s} \sum_{z=0}^{\infty} \left( e^{- \ba}  ( 1 - e^{- \ba})^{z-1} ( 1 - e^{- q}) - e^{\left( \frac{d_1^m P_r - 2z \eta P_s ( (z-1)\ba + q)}{2 \eta P_s } \right)} \left(  e^{ \ba} - 1 \right)^{z-1} \right. \notag \\ & \hspace{6.5cm} \times \left. \left(2 z \eta P_sq - d_1^m P_r + (d_1^m P_r - 2z \eta P_s ) e^{q} \right) E_1\left( \frac{d_1^m P_r + 2\eta P_s \ba}{2 \eta P_s } \right) \right),
\end{align}
\setcounter{equation}{\value{MYtempeqncnt}}
\vspace*{-.4cm}
\end{figure*}
\fi

\noindent where $ \bar{b} \triangleq \frac{\gamma_0 d_2^m \sigma_{n_d}^2}{P_r}$. Note that the factor, $\mathbbm{1}( \gamma_{r,i} < \gamma_o )$, is not part of the outage indicator function, $I_{o,i}$, in \eqref{eq:pout_DF}, because, if the source to relay channel is in outage, i.e., $\gamma_{r,i} < \gamma_o$ or equivalently $\h < \bar{a} \triangleq \frac{\gamma_0 d_1^m \sigma_{n_r}^2}{P_s}$, there is no IT and only EH is carried out for the entire block.



\subsection{Protocol 3: TS-based Protocol for EH and IT with Continuous Time EH in DF Relaying} \label{sec:DFc}

\subsubsection{Description}
In this protocol, similar to Protocol 1, the relay only harvests that much energy within each block that is needed for its transmission. However, in the event of outage at the relay node (i.e., $\h \ge \ba$) the whole block in dedicated for EH. Consequently, in this protocol three types of EH-IT patterns are possible:
\begin{itemize}
\item EH-IT pattern contains only a single block. This is the same as in Protocol 1. For such blocks, $E_i(0) <  \frac{P_r T}{2}$ and the partial block time, $\alpha_i T$ seconds, is used for EH and the remaining block time, $(1-\alpha_i) T $ seconds, is used for IT. This is illustrated in Fig. \ref{fig:DFc} (a).

\item EH-IT pattern contains $n$ consecutive EH blocks due to relay outage, before an EH-IT block. In this case, the energy is accumulated during the EH blocks and only reaches the required level sometime during the EH-IT block. The whole energy is then used for transmission such that $E_o=0$ at the start of the next pattern. This is illustrated in Fig. \ref{fig:DFc} (b) assuming $n=1$.

\item EH-IT pattern contains $n$ consecutive EH blocks due to relay outage, before an IT block. In this case, the energy is accumulated during the EH blocks and exceeds the required level for transmission. Thus the $n$ EH blocks are followed by an IT block and $E_o > 0$ at the start of the next pattern. This is illustrated in Fig. \ref{fig:DFc} (c) assuming $n=1$.

\end{itemize}
\ifCLASSOPTIONpeerreview
\begin{figure}[t]
\centering
     \hspace{-.0cm}\includegraphics[scale=1.15]{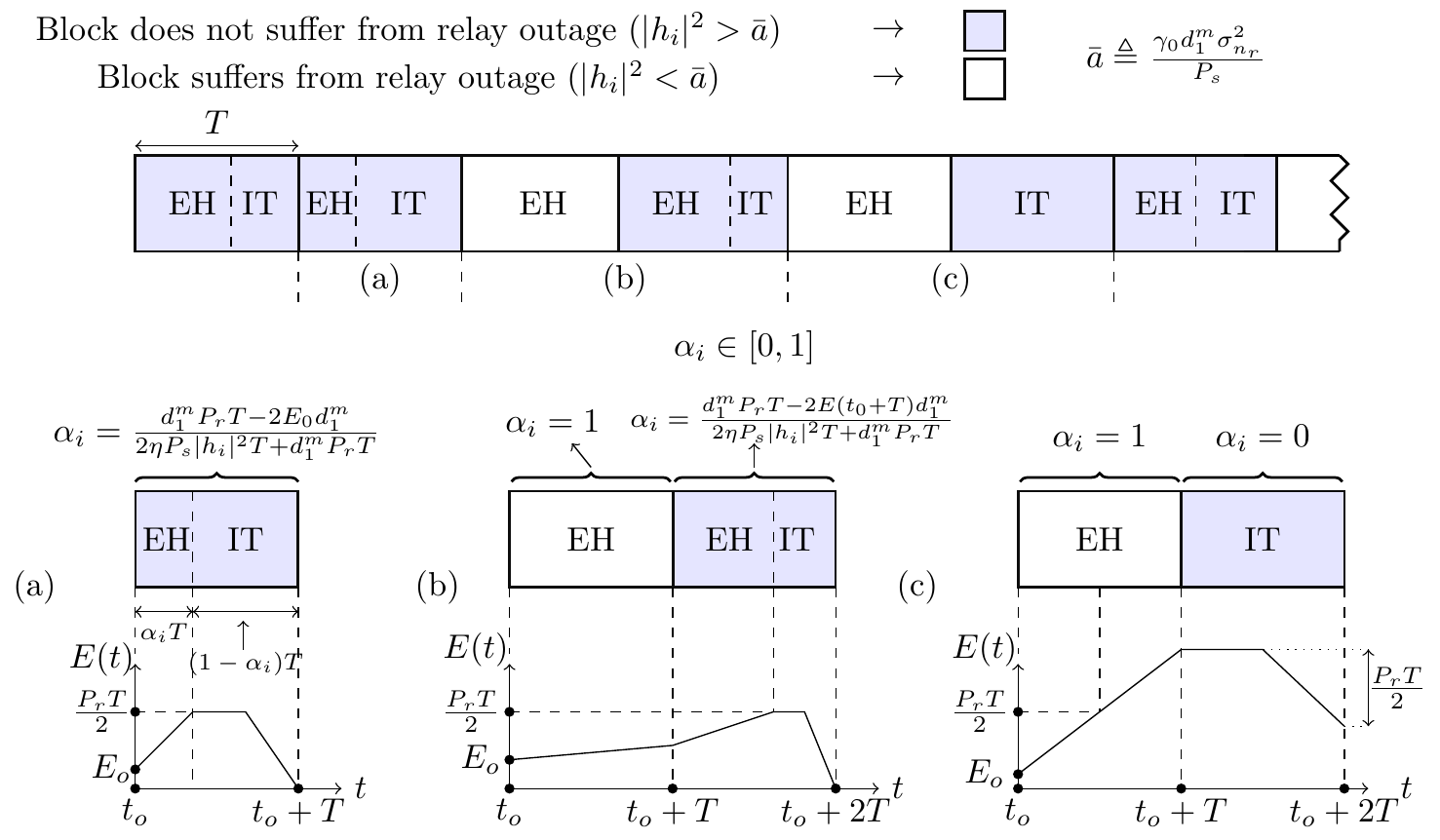}
     \centering
    \caption{Illustration of TS-based protocol for EH and IT in DF relaying, with continuous time EH, $\alpha_i \in [0,1]$ for three different EH-IT patterns (a) single EH-IT block (b) $n=1$ EH block due to relay outage followed by an EH-IT block and (c) $n=1$ EH block due to relay outage followed by an IT block. $n$ is the random number of successive EH blocks due to relay outage, occurring irrespective of accumulated energy, in continuous time EH (DF only).}
    \label{fig:DFc}
\end{figure}
\else
\fi

\subsubsection{Energy Analysis}

We know that $\alpha_i = 0$ for an IT block and $\alpha_i = 1$ for an EH block. The value of $\alpha_i$ for an EH-IT block can be determined as follows. The harvested energy at the time instant, $\alpha_i T$, during the $i$th EH-IT block is given by
\begin{align}\label{eq:EH1_DFc}
      E_i(\alpha_i T) =  \frac{\eta P_s |h_i|^2}{d_1^m} \alpha_i T + E_i(0).
\end{align}

Note that the term $E_i(0)$ was not present in \eqref{eq:EH1_AFc} for AF relaying because $E_i(0)$ was $0$. Similar to Protocol 1, the harvested energy at the time, $\alpha_i T$ should be equal to $P_r \frac{(1-\alpha_i) T}{2}$ (see \eqref{eq:EH2_AFc}). The value of $\alpha_i$ for EH-IT block can be obtained by equating \eqref{eq:EH1_DFc} to $P_r \frac{(1-\alpha_i) T}{2}$, i.e.,
\begin{align}\label{eq:alpha_t_DFc}
     \alpha_i = \frac{d_1^m P_r T - 2 E_i(0) d_1^m}{2\eta P_s | h_i|^2 T  + d_1^m P_r T },
\end{align}

Thus, $\alpha_i$ and $E_i(T)$ for any general $i$th block are given by
%
\ifCLASSOPTIONpeerreview
\begin{align}\label{eq:alpha_t_DFc}
      \alpha_i = \begin{cases}
                    \frac{d_1^m P_r T - 2 E_i(0) d_1^m}{2\eta P_s | h_i|^2 T  + d_1^m P_r T }, & \h \ge \ba \;,\; E_i(0) < \frac{P_r T}{2} \\
                    0 , & \h \ge \ba \;,\;  E_i(0) > \frac{P_r T}{2} \\
                    1 , & \h < \ba \end{cases}
\end{align}
\begin{align}\label{eq:Ei_DFc}
                      E_i(T) = \begin{cases}
                    0, &  0 < \alpha_i < 1 \\
                    E_i(0) + \frac{\eta P_s |h_i|^2 T}{d_1^m}, & \alpha_i=1 \\
                    E_i(0) - \frac{P_r T}{2} , & \alpha_i=0
       \end{cases}
\end{align}
\else
\begin{align}\label{eq:alpha_t_DFc}
      \alpha_i = \begin{cases}
                    \frac{d_1^m P_r T - 2 E_i(0) d_1^m}{2\eta P_s | h_i|^2 T  + d_1^m P_r T }, & \h \ge \ba , E_i(0) < \frac{P_r T}{2} \\
                    0 , & \h \ge \ba ,  E_i(0) > \frac{P_r T}{2} \\
                    1 , & \h < \ba \end{cases}
\end{align}
\begin{align}\label{eq:Ei_DFc}
                      E_i(T) = \begin{cases}
                    0, &  0 < \alpha_i < 1 \\
                    E_i(0) + \frac{\eta P_s |h_i|^2 T}{d_1^m}, & \alpha_i=1 \\
                    E_i(0) - \frac{P_r T}{2} , & \alpha_i=0
       \end{cases}
\end{align}
\fi

Using \eqref{eq:alpha_t_DFc}, the block throughput $\tau_i$ for the $i$th block is given by
\begin{align}\label{eq:tau_DF}
      \tau_i = \frac{(1 - I_{o,i}) (1 - \alpha_i)}{2},
\end{align}
where $\alpha_i$ and $I_{o,i}$ are given in \eqref{eq:alpha_t_DFc} and \eqref{eq:pout_DF}, respectively. Note that if there is an outage at the relay node, $\alpha_i$ is set to $1$, and the block throughput $\tau_i$ will be zero.

\subsubsection{Throughput Analysis}\label{sec:ana_tau_DFc}
It does not seem tractable to find the exact throughput for Protocol 3. However, it is possible to derive a lower bound on the throughput by assuming that $E_o=0$. As illustrated in Fig.~\ref{fig:DFc}, only the EH-IT pattern in Fig.~\ref{fig:DFc} (c) can lead to extra harvested energy and consequently $E_o$ being greater than 0. However, if the relay is in outage, it means that the channel quality is poor and the probability that the amount of energy harvested from $n$ successive blocks is greater than $\frac{P_r T}{2}$ is very low. Moreover, the occurrence of successive outages is also less probable. Hence, we assume that if the EH-IT pattern in Fig.~\ref{fig:DFc} (c) occurs, it results in $E_o=0$. Using this assumption, we derive a lower bound on the throughput for Protocol 3, which is given in Theorem~\ref{Th3} below.

\begin{theorem}\label{Th3}
A lower bound on the throughput for the TS-based Protocol for EH and IT with continuous time EH in DF relaying is given by
\ifCLASSOPTIONpeerreview
\begin{align}\label{eq:tau_DFc_ana}
\tau  &\ge   \frac{e^{-(\ba+\bb)}}{4 \eta P_s} \sum_{n=0}^{\infty} \left( e^{- \ba}  ( 1 - e^{- \ba})^{n-1} ( 1 - e^{- q}) - e^{\left( \frac{d_1^m P_r - 2n \eta P_s ( (n-1)\ba + q)}{2 \eta P_s } \right)} \left(  e^{ \ba} - 1 \right)^{n-1} \right. \notag \\ & \hspace{5cm} \times \left. \left(2 n \eta P_sq - d_1^m P_r + (d_1^m P_r - 2n \eta P_s ) e^{q} \right) E_1\left( \frac{d_1^m P_r + 2\eta P_s \ba}{2 \eta P_s } \right) \right),
\end{align}
\else
\addtocounter{equation}{1}
\eqref{eq:tau_DFc_ana} at the bottom of the page,
\fi
where $E_1(x) = \int_{x}^{\infty} \frac{e^{-t}}{t} dt$ is the exponential integral, $\ba$ and $\bb$, are defined below \eqref{eq:pout_DF}, $q \triangleq \min \left( \ba, \frac{d_1^m P_r}{2 \eta P_s n} \right)$ and $n$ denotes the number of successive EH blocks due to relay outage.
\end{theorem}

\begin{IEEEproof}
See Appendix \ref{app:C}.
\end{IEEEproof}

\begin{remark}\label{R1}
The factor inside the summation in \eqref{eq:tau_DFc_ana} decays very quickly to $0$ for moderate to large $n$ (approx. $n \ge 10$).  This is because the probability that $n$ successive blocks suffer from relay outage decays very quickly to $0$ for large $n$. Thus, we can accurately evaluate the infinite summation in \eqref{eq:tau_DFc_ana} using $n=10$ terms only. The simulation results in Section \ref{sec:sim} show that even with $n=10$, the analytical throughput in \eqref{eq:tau_DFc_ana} is an accurate lower bound on the actual throughput calculated using simulations.
\end{remark}

\subsection{Protocol 4: TS-based Protocol for EH and IT with Discrete Time EH in DF Relaying}  \label{sec:DFd}

\ifCLASSOPTIONpeerreview
\begin{figure}[t]
\centering
     \hspace{-.0cm}\includegraphics[scale=1.2]{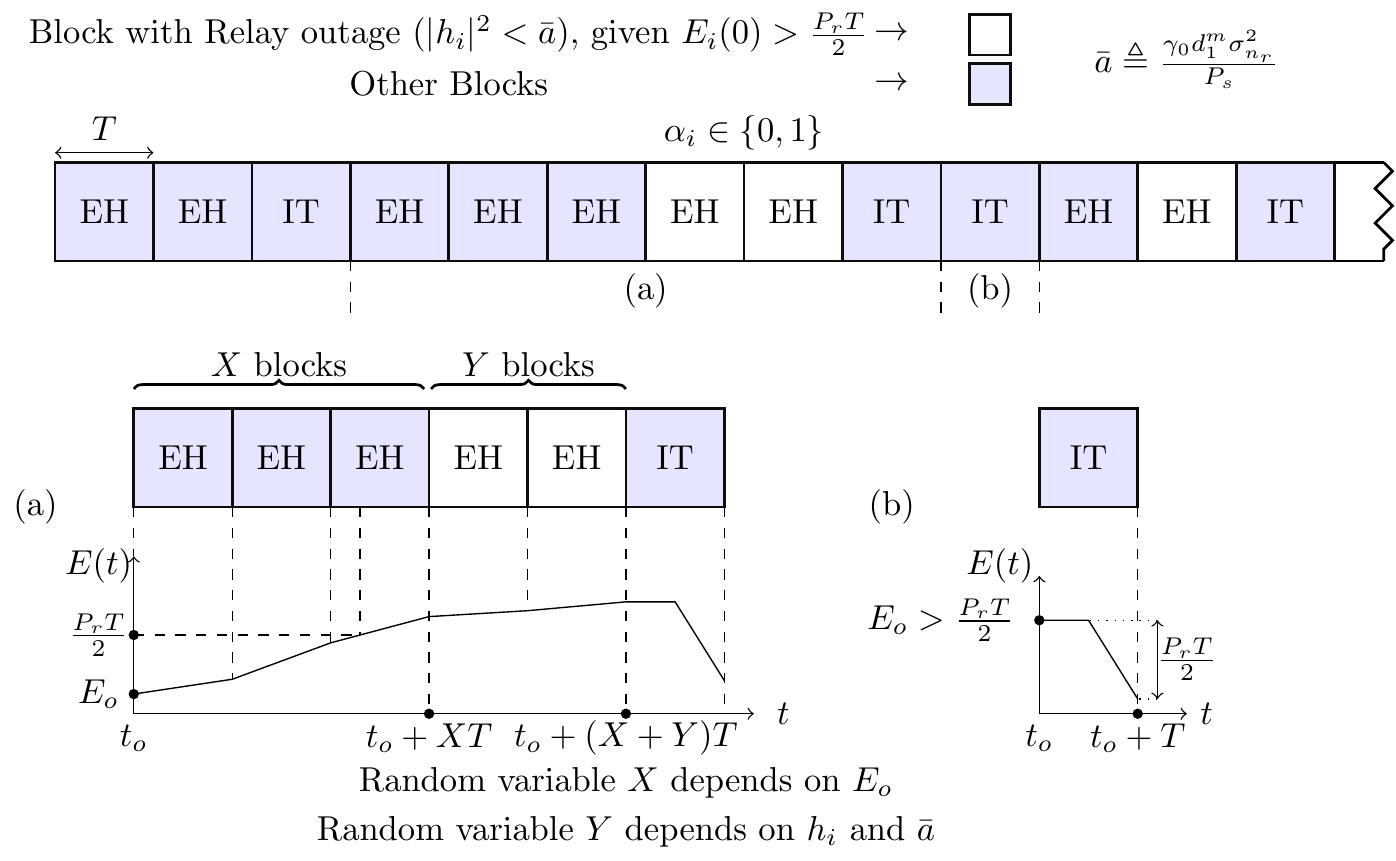}
     \centering
    \caption{Illustration of TS-based protocol for EH and IT in DF relaying with discrete time EH, i.e., $\alpha_i \in \{0,1\}$, for two different EH-IT patterns (a) $X=3$ and $Y=2$ EH blocks before an IT block and (b) $X=0$ and $Y=0$ (single IT block).}
    \label{fig:DFd}
\end{figure}
\else
\fi

\ifCLASSOPTIONpeerreview
\else
\begin{figure*}[t]
\centering
     \hspace{-.0cm}\includegraphics[scale=1.0]{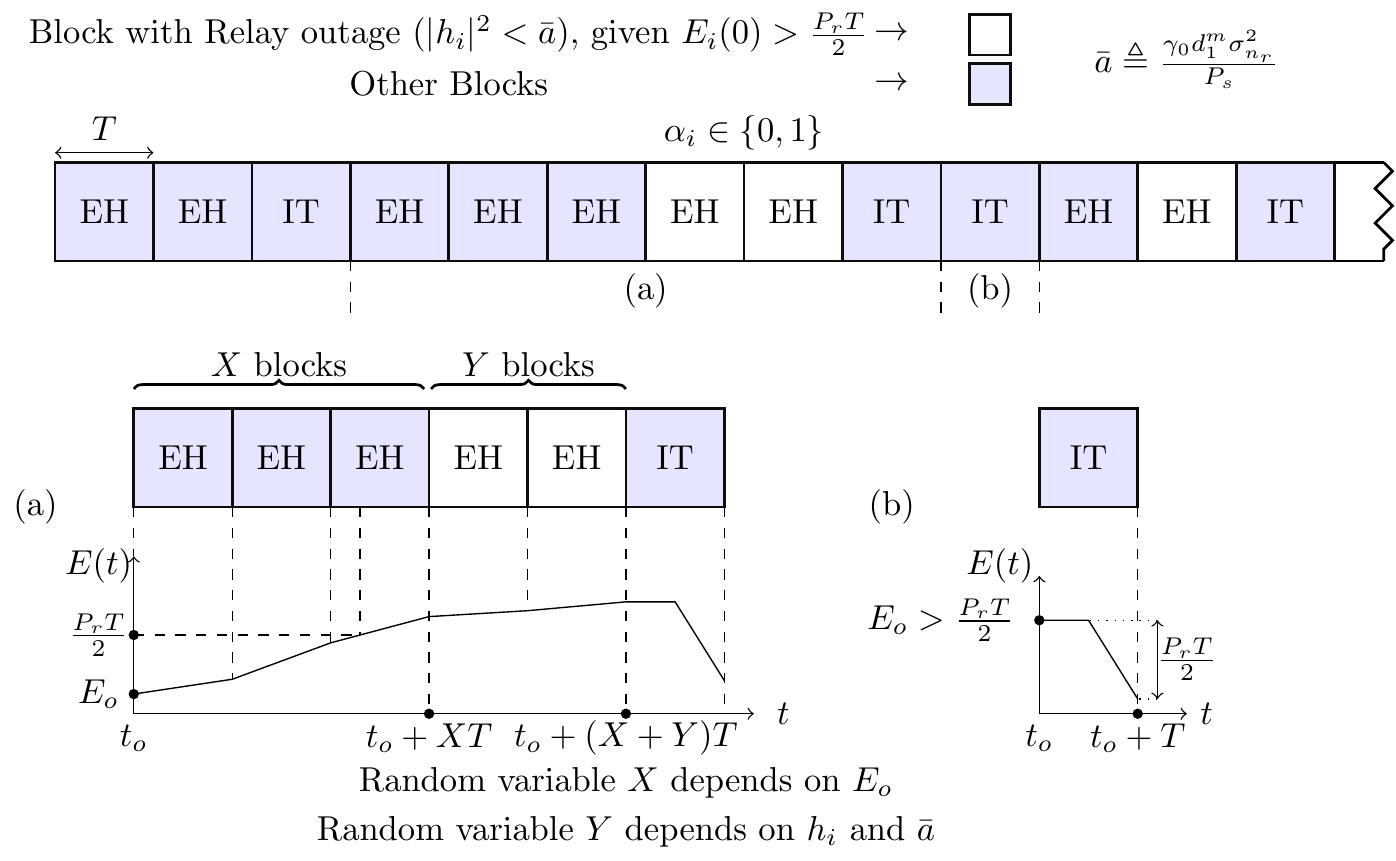}
     \centering
    \caption{Illustration of TS-based protocol for EH and IT in DF relaying with discrete time EH, i.e., $\alpha_i \in \{0,1\}$, for two different EH-IT patterns (a) $X=3$ and $Y=2$ EH blocks before an IT block and (b) $X=0$ and $Y=0$ (single IT block).}
    \label{fig:DFd}
\end{figure*}
\fi

\subsubsection{Description}
In this protocol, each block is dedicated either for EH ($\alpha_i = 1$) or IT ($\alpha_i = 0$). Similar to Protocol 2, the relay checks its available energy at the start of each block. It continues to harvest energy until the harvested energy becomes greater than or equal to the desired energy level, $\frac{P_r T}{2}$. Different from Protocol 2, after the desired energy, $\frac{P_r T}{2}$, is achieved, the relay executes IT only if the block does not suffer from relay outage. In case of a relay outage, the relay continues to perform EH. Consequently, in this protocol the following two types of EH-IT patterns are possible:
\begin{itemize}
\item EH-IT pattern contains $X$ consecutive EH blocks until desired energy is achieved, followed by $Y$ consecutive EH blocks due to relay outage which prevent IT, before an IT block. This is illustrated in  Fig.~\ref{fig:DFd} (a) with $X=3$ and $Y=2$, where the white-shaded blocks represent the EH blocks due to relay outage.

\item EH-IT pattern contains a single IT block because $X=0$ and $Y=0$. This is illustrated in  Fig.~\ref{fig:DFd} (b).
\end{itemize}

The random variable $X$ here is defined the same as in Protocol 2. However, the random variable $Y$ depends on the quality of source-to-relay channel, $h_i$, and is different from RV $n$ (for Protocol 3). This is because $Y$ is the random number of successive EH blocks due to relay outage, occurring after sufficient amount of energy has been harvested for IT in Protocol 4. Whereas, $n$ is the random number of successive EH blocks due to relay outage, occurring irrespective of the accumulated energy, in Protocol 3.

\subsubsection{Energy Analysis} Using the same steps as before, the value of $\alpha_i$ and $E_i(T)$ for the $i$th block can be expressed as

\begin{align}\label{eq:alpha_t_DFd}
      \alpha_i = \begin{cases}
                    1 , & E_i(0) < \frac{P_r T}{2} \\
                    1 , & E_i(0) > \frac{P_r T}{2} \;,\; \h < \ba \\
                    0 , & E_i(0) > \frac{P_r T}{2} \;,\; \h \ge \ba
       \end{cases}
\end{align}
\begin{align}\label{eq:Ei_DFd}
        E_i(T) = \begin{cases}
                    E_i(0) + \frac{\eta P_s |h_i|^2}{d_1^m} T, & \alpha_i=1 \\
                    E_i(0) - \frac{P_r T}{2} , & \alpha_i=0
       \end{cases}
\end{align}

\subsubsection{Throughput Analysis}\label{sec:DFd_ana}

The block throughput $\tau_i$ for the $i$th block can be calculated using \eqref{eq:tau_DF}, where $\alpha_i$ is given by \eqref{eq:alpha_t_DFd}. This requires the determination of the PMF of $Y$, the PDF of $E_o$, and the conditional PMF of $\bar{X} \triangleq X-1$, given the value of $E_o$. The latter is given in Lemma~\ref{L2} already. The PMF of $Y$ is given in the lemma below.
\begin{lemma}\label{L3}
The PMF of $Y$ is given by
\begin{align}\label{eq:PMF_Y}
p_{Y}(y) = \left(1-p_{o,r} \right) p_{o,r}^y,
\end{align}
where relay outage probability, $p_{o,r} = 1 - e^{-\ba}$, and $y = \{0,1,2,\hdots\}$.
\end{lemma}
\begin{IEEEproof}
After sufficient amount of energy, $\frac{P_r T}{2}$, has been harvested for IT, the probability of the discrete random variable $Y$ being $y$ is the probability that successive $y$ blocks, before an IT block, suffer from relay outage. Since, the $i$th block suffers from relay outage if $\h < \ba$, the expected relay outage probability, $p_{o,r}$ is obtained by finding the expectation of the indicator function, $\mathbbm{1} (\h < \ba)$, over $h_i$, i.e., $p_{o,r} = \mathbb{E}_{h_i} \left\{ \mathbbm{1} (\h < \ba) \right\} = 1 - e^{-\ba}$, where $\ba > 0$. Since, the occurrence of the relay outage in any one block is independent from the other, PMF of $Y$ is given by $p_{Y}(y) = \left(1-p_{o,r} \right) p_{o,r}^y$.
\end{IEEEproof}

\ifCLASSOPTIONpeerreview
\else

\begin{figure*}[t]
\centering
    \begin{minipage}[h]{0.49\textwidth}
    \centering
    \includegraphics[width=1.0 \textwidth]{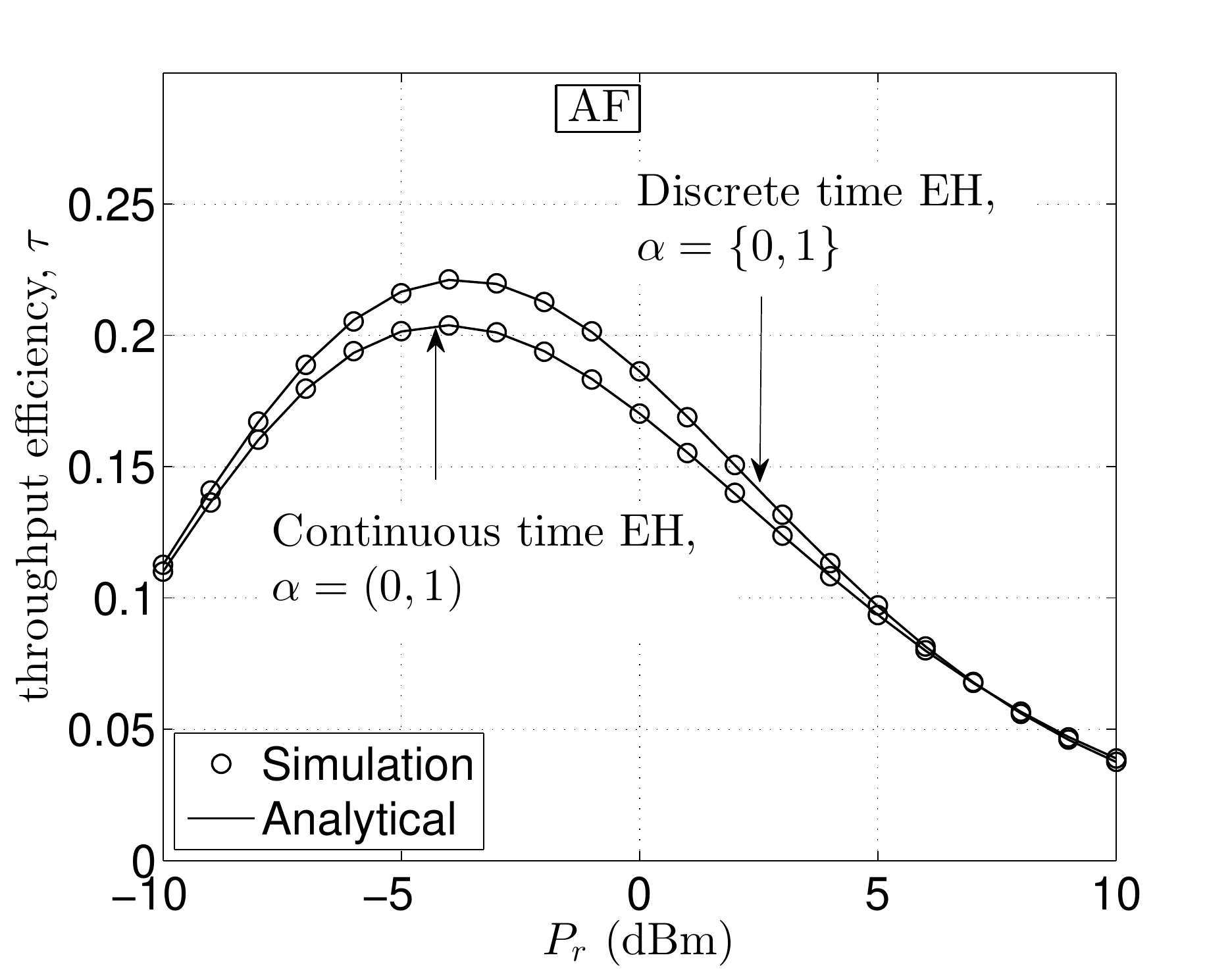}
  \caption{Throughput efficiency, $\tau$ with respect to the preset relay power, $P_r$, for TS-based protocols for EH and IT in AF relaying with continuous time and discrete time EH.}
  \label{fig:Fig1}
  \end{minipage}
    \centering
    \begin{minipage}[h]{0.49\textwidth}
    \centering
    \includegraphics[width=1.0 \textwidth]{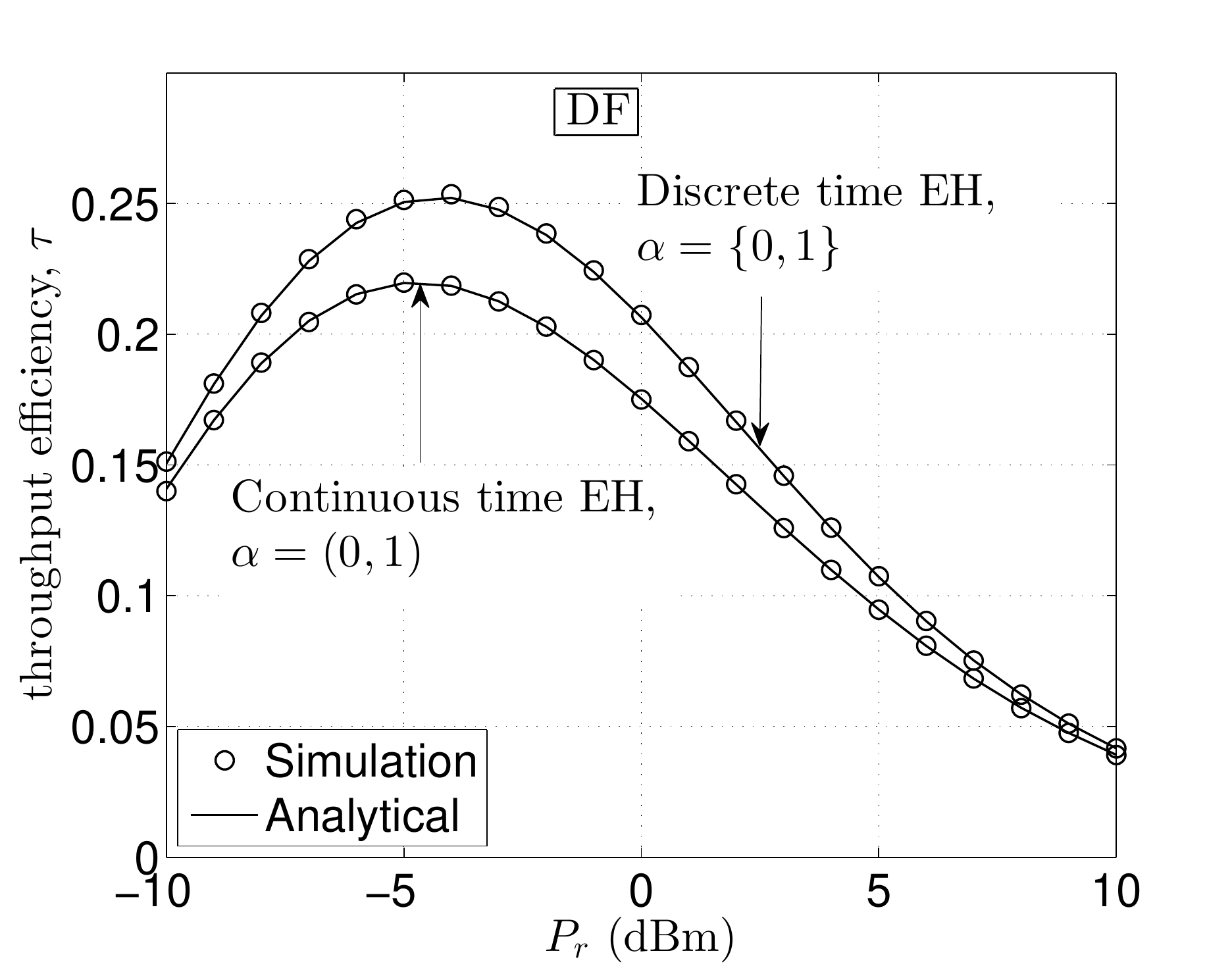}
  \caption{Throughput efficiency, $\tau$ with respect to the preset relay power, $P_r$, for TS-based protocols for EH and IT in AF relaying with continuous time and discrete time EH.}
  \label{fig:Fig2}
  \end{minipage}
\end{figure*}

\fi

It does not seem tractable to determine the exact distribution of $E_o$ for this protocol. Compared to Protocol 2, in Protocol 4 we have extra harvested energy in the additional $Y$ EH blocks due to relay outage. However, if the relay is in outage, it means that the channel quality is poor and for analysis we can neglect the harvested energy from $Y$ EH blocks. Using this assumption, the PDF of $E_o$ is the same as given in Lemma \ref{L1}. Thus, using Lemmas~\ref{L1}, \ref{L2} and \ref{L3}, we can derive a lower bound on the throughput, which is given in Theorem~\ref{Th4} below.
%
%
\begin{theorem}\label{Th4}
A lower bound on the throughput for the TS-based Protocol for EH and IT with discrete time EH in DF relaying is given by
\begin{align}\label{eq:tau_DFd_ana}
\tau \ge \frac{\eta P_s e^{-\left( \ba + \bb \right)} }{P_r d_1^m e^{-\ba} + 2 \eta P_s}.
\end{align}
\end{theorem}
\begin{IEEEproof}
See Appendix \ref{app:D}.
\end{IEEEproof}

\begin{remark}\label{R2}
The simulation results in Section \ref{sec:sim} show that the analytical throughput in \eqref{eq:tau_DFd_ana} is an accurate lower bound on the actual throughput calculated using simulations.
\end{remark}

\section{Numerical Results and Discussion}\label{sec:sim}

In this section, we present numerical results to demonstrate the performance of the proposed protocols as a function of the system parameters. We adopt the realistic simulation parameter values as suggested in recent survey papers on energy harvesting~\cite{Lu-14,Gunduz-14,Huang-15-A,Visser-13-A,Tabassum-15}. We set the source transmission power, $P_s = 46$ dBm, and the path loss exponent, $m = 3$. Since wireless energy harvesting is practical over shorter distances, we set the distances, $d_1 = 35$ and $d_2 = 10$ meters. The state-of-the-art circuit design establishes that RF signals over a wide range of frequencies can be rectified at an efficiency higher than $50\%$ \cite{Lu-14}. Hence, we set the energy conversion efficiency to $\eta = 0.5$. Unless otherwise stated, we set the threshold SNR, $\gamma_o = 60$ dB, and the noise variances at the relay and the destination nodes, $\sigr = -70$ dBm and $\sigd = -100$ dBm respectively. Note that we choose a larger value of $\gamma_o$ and $\sigr$ to allow the relay outage to impact the throughput results in Theorems 3 and 4. Note that later in this section, in order to investigate the system throughput, we will consider a larger range of values of $\gamma_o$, $\sigr$ and $\sigd$.


\subsection{Verification of Analytical Results}

In this subsection, we present simulation results to verify the analytical results in Theorems \ref{Th1}-\ref{Th4}. In the simulations, the throughput is evaluated by averaging out the block throughput $\tau_i$ over a 100,000 blocks, while generating independent fading channels, $h_i$ and $g_i$, for each block.

Figs. \ref{fig:Fig1} and \ref{fig:Fig2} plot the analytical and simulation based throughput, $\tau$ versus relay power, $P_r$, for Protocols 1-4.\footnote{Since the energy harvesting time $\alpha_i$ is different for each block, we cannot plot throughput versus energy harvesting time for Protocols 1-4. The tradeoff between the throughput $\tau$ and energy harvesting time can be found in \cite{Nasir-13-A}, where the energy harvesting time is fixed.} In Fig. \ref{fig:Fig1}, the analytical results for AF relaying are plotted by numerically evaluating the throughput in Theorems \ref{Th1} and \ref{Th2}. The simulation results match perfectly with the analytical results. This is to be expected since Theorems \ref{Th1} and \ref{Th2} are exact results. In Fig. \ref{fig:Fig2} the analytical lower bounds for DF relaying are plotted by using $n=10$ summation terms in Theorem \ref{Th3} and by numerically evaluating the throughput in Theorem \ref{Th4}. We can see that the analytical results are a tight lower bound on the actual throughput. Note that decreasing $\sigr$ further from $-70$ dBm decreases the relay outage making the lower bound even more tight (the results are not included in the figure for the sake of clarity). This validates the results in Theorems \ref{Th3}-\ref{Th4}.

It can be seen from Figs. \ref{fig:Fig1} and \ref{fig:Fig2} that the throughput can vary significantly with $P_r$. In order to achieve the maximum throughput, we have to choose the optimal relay transmission power. Given the complexity of the expressions in Theorems \ref{Th1}-\ref{Th4}, it seems intractable to find the closed-form expressions for the optimal relay power, $P_r$, which would maximize the throughput, $\tau$. However, such an optimization can be done offline for the given system parameters. In the following subsections, we adopt the maximum throughput for some optimal preset relay power $P_r$ as the figure of merit and refer to it as the \emph{optimal throughput}.

\ifCLASSOPTIONpeerreview
\begin{figure}[t]
\centering
    \begin{minipage}[h]{0.48\textwidth}
    \centering
    \includegraphics[width=1.05 \textwidth]{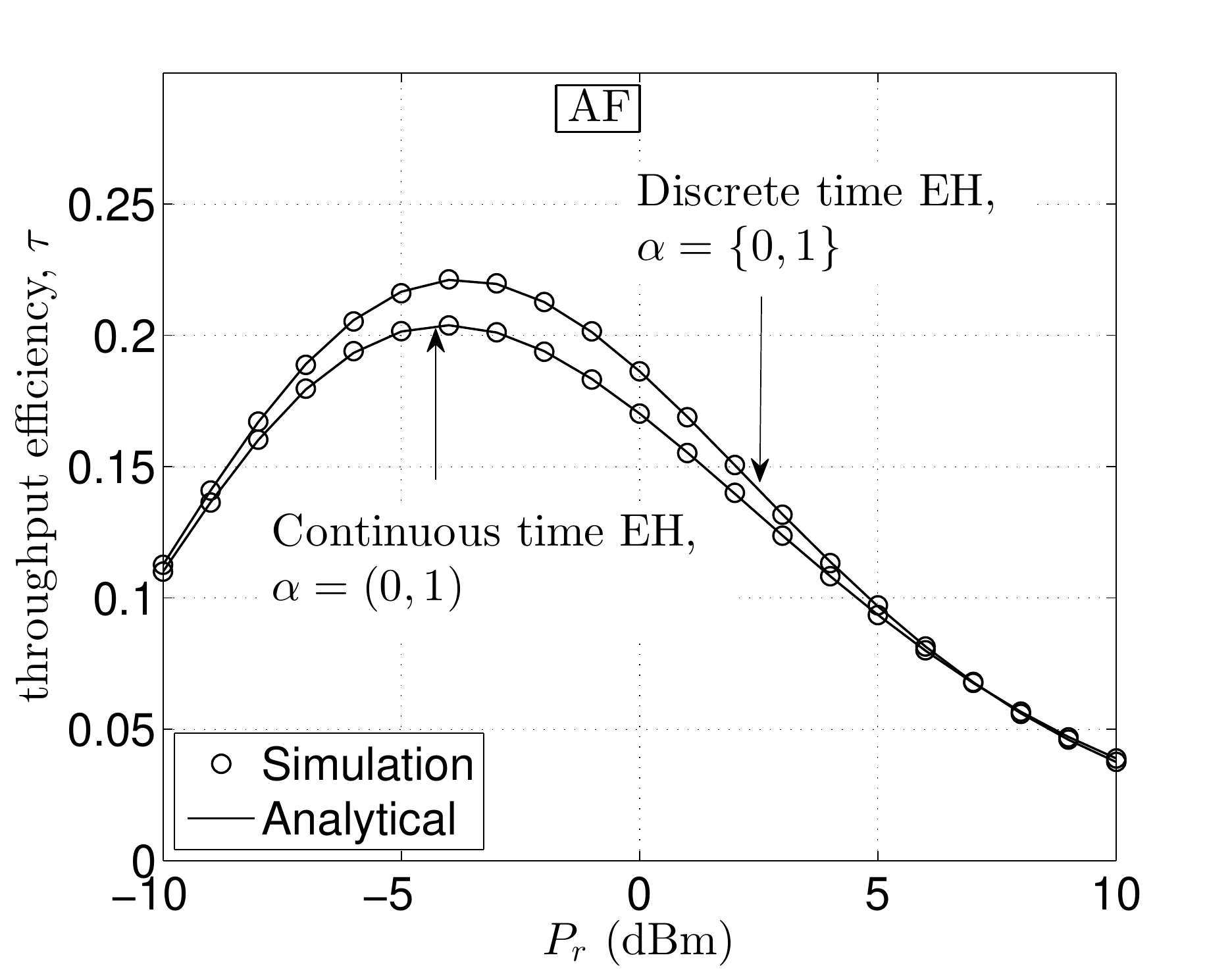}
  \caption{Throughput efficiency, $\tau$ with respect to the preset relay power, $P_r$, for TS-based protocols for EH and IT in AF relaying with continuous time and discrete time EH.}
  \label{fig:Fig1}
  \end{minipage}
  \hspace{0.2cm}
    \centering
    \begin{minipage}[h]{0.48\textwidth}
    \centering
    \includegraphics[width=1.05 \textwidth]{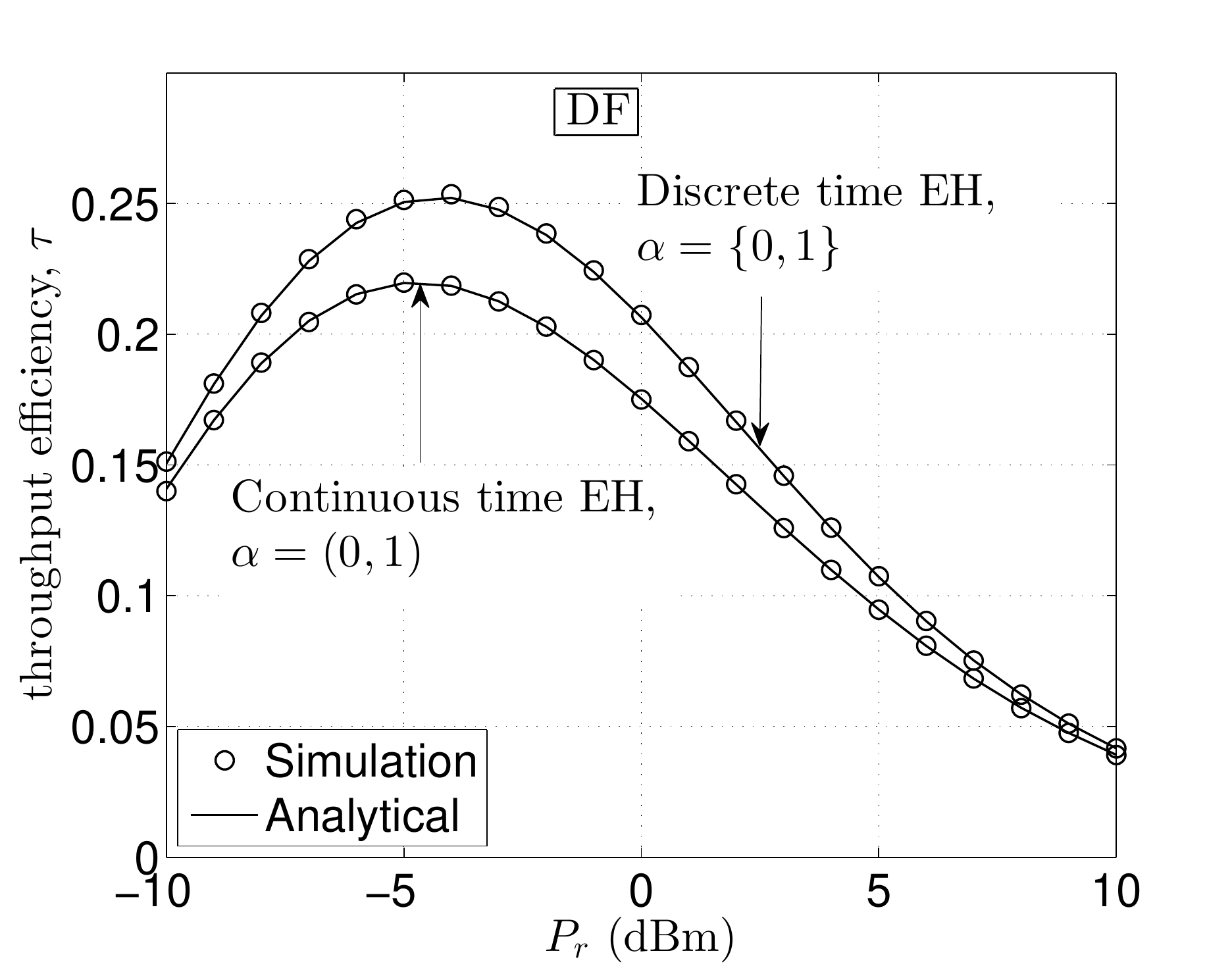}
  \caption{Throughput efficiency, $\tau$ with respect to the preset relay power, $P_r$, for TS-based protocols for EH and IT in AF relaying with continuous time and discrete time EH.}
  \label{fig:Fig2}
  \end{minipage}
\end{figure}

\else
\fi

\ifCLASSOPTIONpeerreview
\else
\begin{figure*}[t]
    \centering
    \begin{minipage}[h]{0.48\textwidth}
    \centering
    \includegraphics[width=1.05 \textwidth]{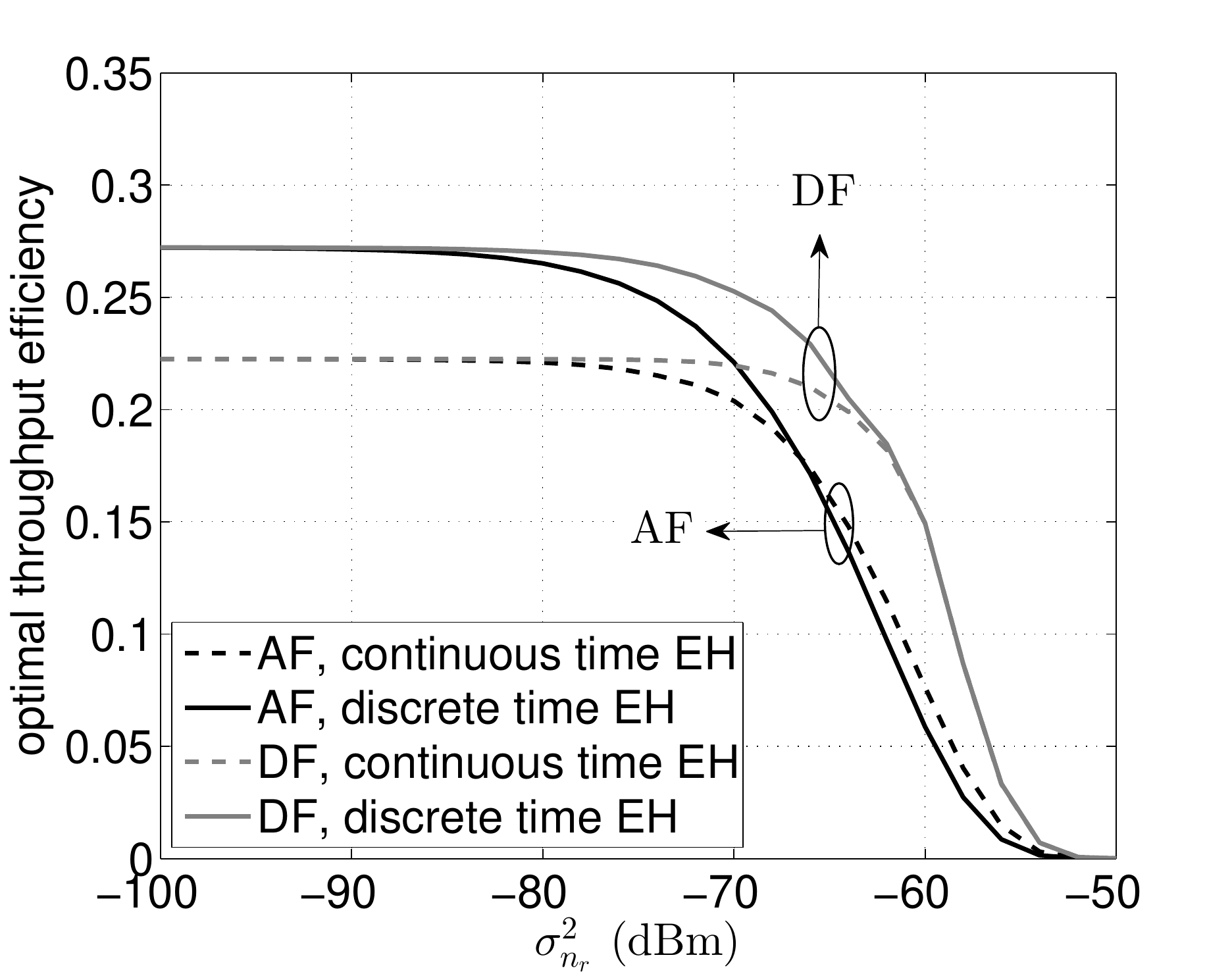}
  \caption{Optimal throughput efficiency versus relay noise variance, $\sigr$, for TS-based protocols for EH and IT in AF and DF relaying with continuous time and discrete time EH.}
  \label{fig:Fig3}
  \end{minipage}
    \hspace{0.2cm}
    \begin{minipage}[h]{0.48\textwidth}
    \centering
    \includegraphics[width=1.05 \textwidth]{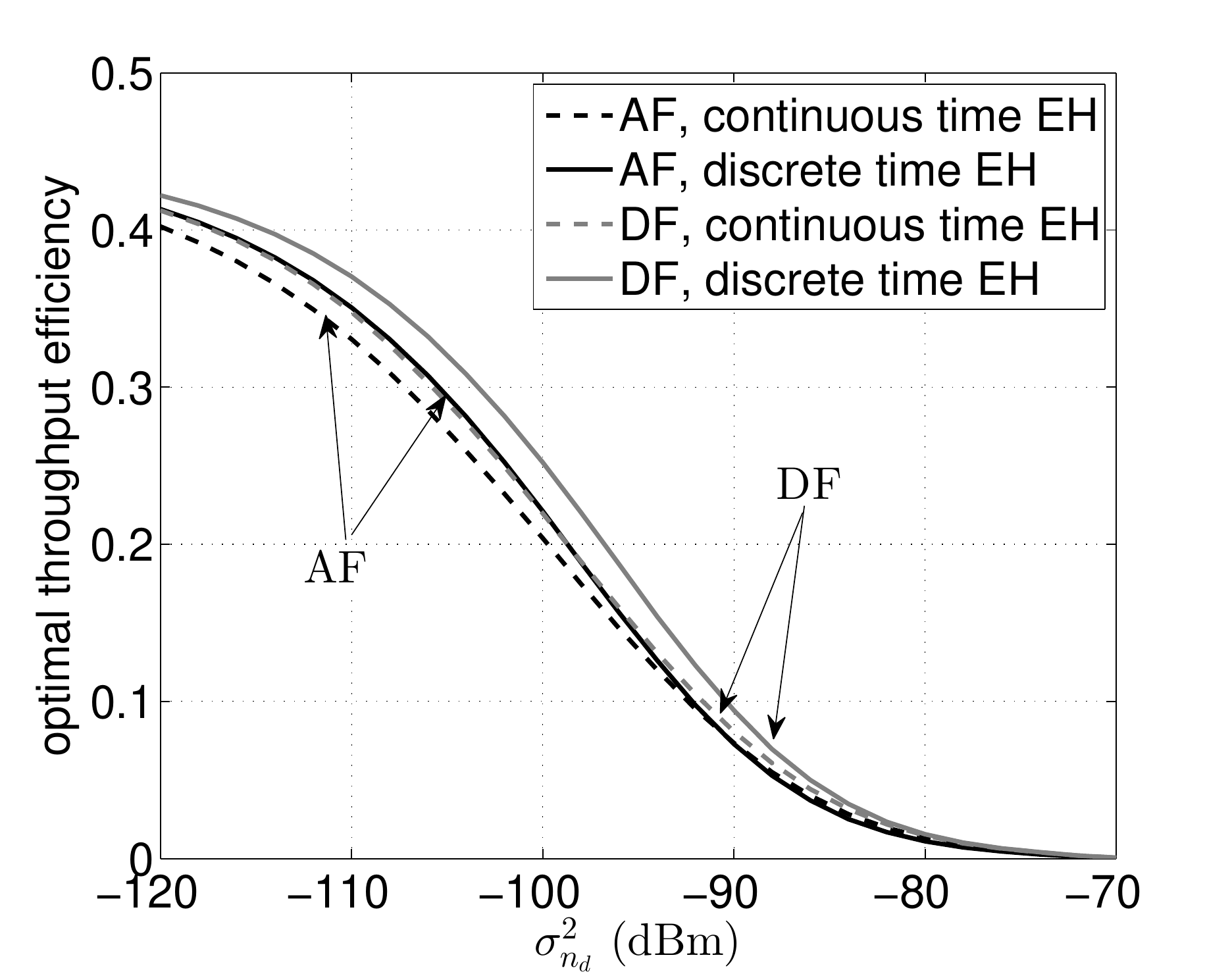}
  \caption{Optimal throughput efficiency versus destination noise variance, $\sigd$, for TS-based protocols for EH and IT in AF and DF relaying with continuous time and discrete time EH.}
  \label{fig:Fig4}
  \end{minipage}
\end{figure*}
\fi

\subsection{Performance of the Proposed Protocols}

Figs. \ref{fig:Fig3} and \ref{fig:Fig4} plot the optimal throughput for Protocols 1-4 as a function of the relay and the destination noise variances, $\sigr$ and $\sigd$, respectively. We can see from Figs. \ref{fig:Fig3} and \ref{fig:Fig4} that at very low relay noise variance, $\sigr$, or very high destination noise variance, $\sigd$, the performance of the protocols for AF and DF relaying is the same. As the relay noise variance increases or the destination noise variance decreases, the protocols for DF relaying outperform AF relaying. These trend are generally inline with the performance of AF and DF relaying in traditional non-energy harvesting relaying systems~\cite{Shrestha-08-P}.

Figs. \ref{fig:Fig3} and \ref{fig:Fig4} also show that for AF relaying, continuous time EH protocols outperform discrete time EH protocols, except for very low relay noise variance or very low destination noise variance. For DF relaying, the performance of the continuous time EH and discrete time EH protocols is the same except for very low relay noise variance or very low destination noise variance. This can be intuitively explained as follows. A lower noise variance decreases the outage probability and the discrete time EH protocols benefit more from this by correct IT during the whole block time, compared to the continuous time EH protocols. Conversely, for higher noise variance, the outage probability increases. Thus, the performance of discrete time EH protocols suffers more by erroneous IT leading to a waste of resources during the whole block time.

\ifCLASSOPTIONpeerreview
\begin{figure*}[t]
    \centering
    \begin{minipage}[h]{0.47\textwidth}
    \centering
    \includegraphics[width=1.01 \textwidth]{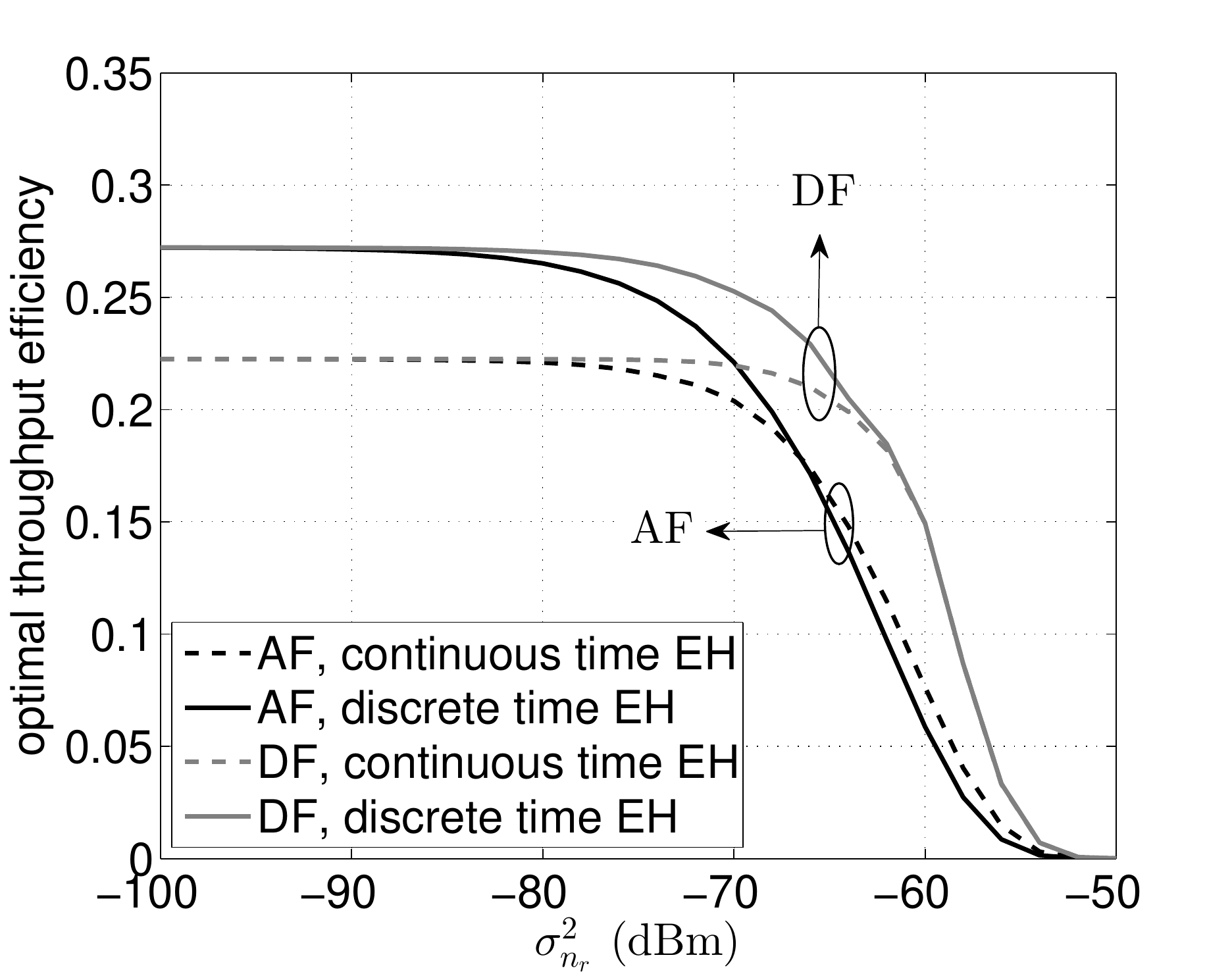}
  \caption{Optimal throughput efficiency versus relay noise variance, $\sigr$, for TS-based protocols for EH and IT in AF and DF relaying with continuous time and discrete time EH.}
  \label{fig:Fig3}
  \end{minipage}
    \hspace{0.3cm}
    \begin{minipage}[h]{0.47\textwidth}
    \centering
    \includegraphics[width=1.01 \textwidth]{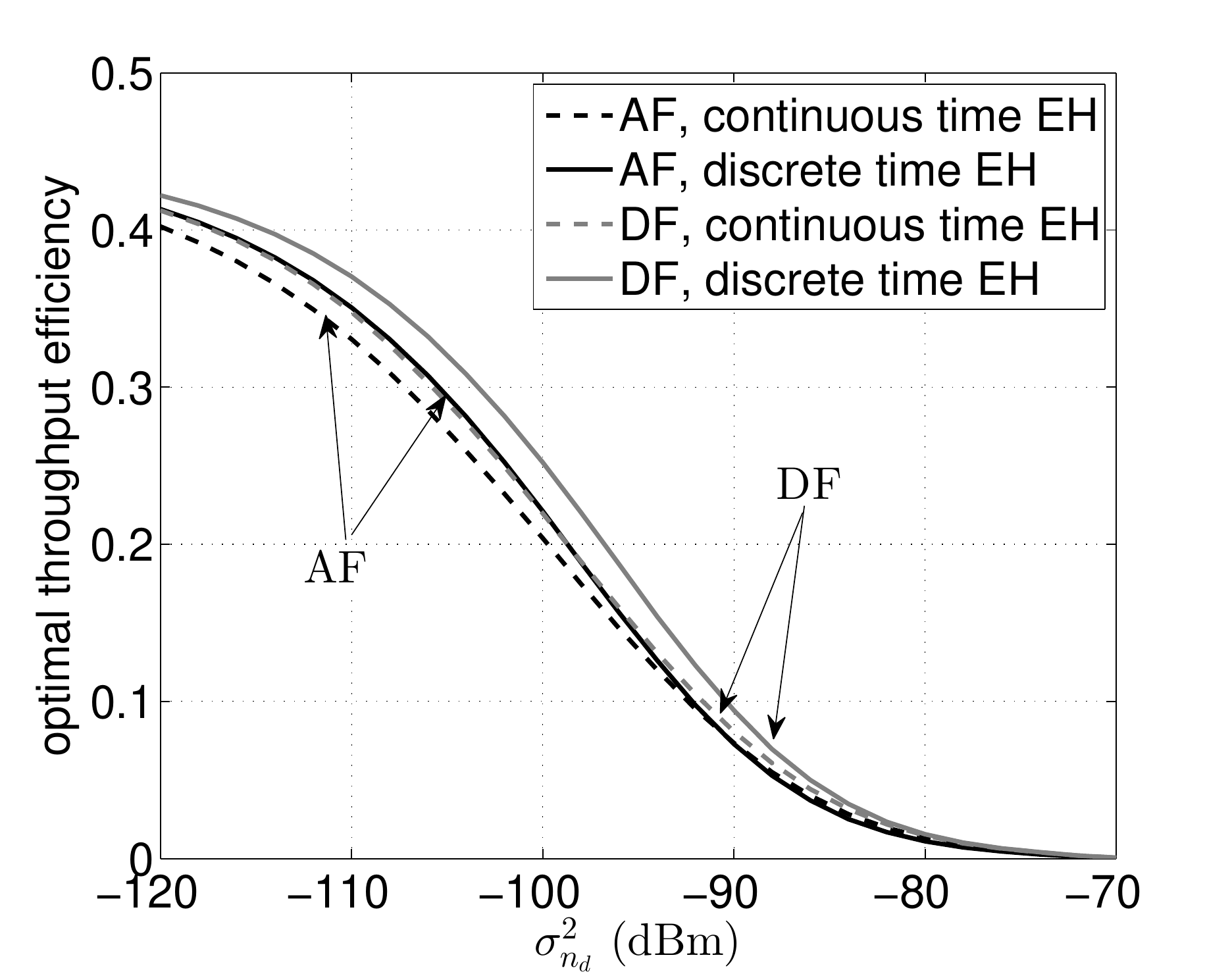}
  \caption{Optimal throughput efficiency versus destination noise variance, $\sigd$, for TS-based protocols for EH and IT in AF and DF relaying with continuous time and discrete time EH.}
  \label{fig:Fig4}
  \end{minipage}
\end{figure*}
\else
\fi

\subsection{Comparison with Existing TS-based Fixed Time-duration EH Protocol}

Finally, we compare the performance of the proposed continuous and discrete time EH protocols and the fixed time-duration EH protocol in \cite{Nasir-13-A} for AF relaying. The protocol in \cite{Nasir-13-A} considers fixed time-duration EH and does not allow relay energy accumulation.

\begin{figure}[t]
    \centering
    \includegraphics[width=0.49 \textwidth]{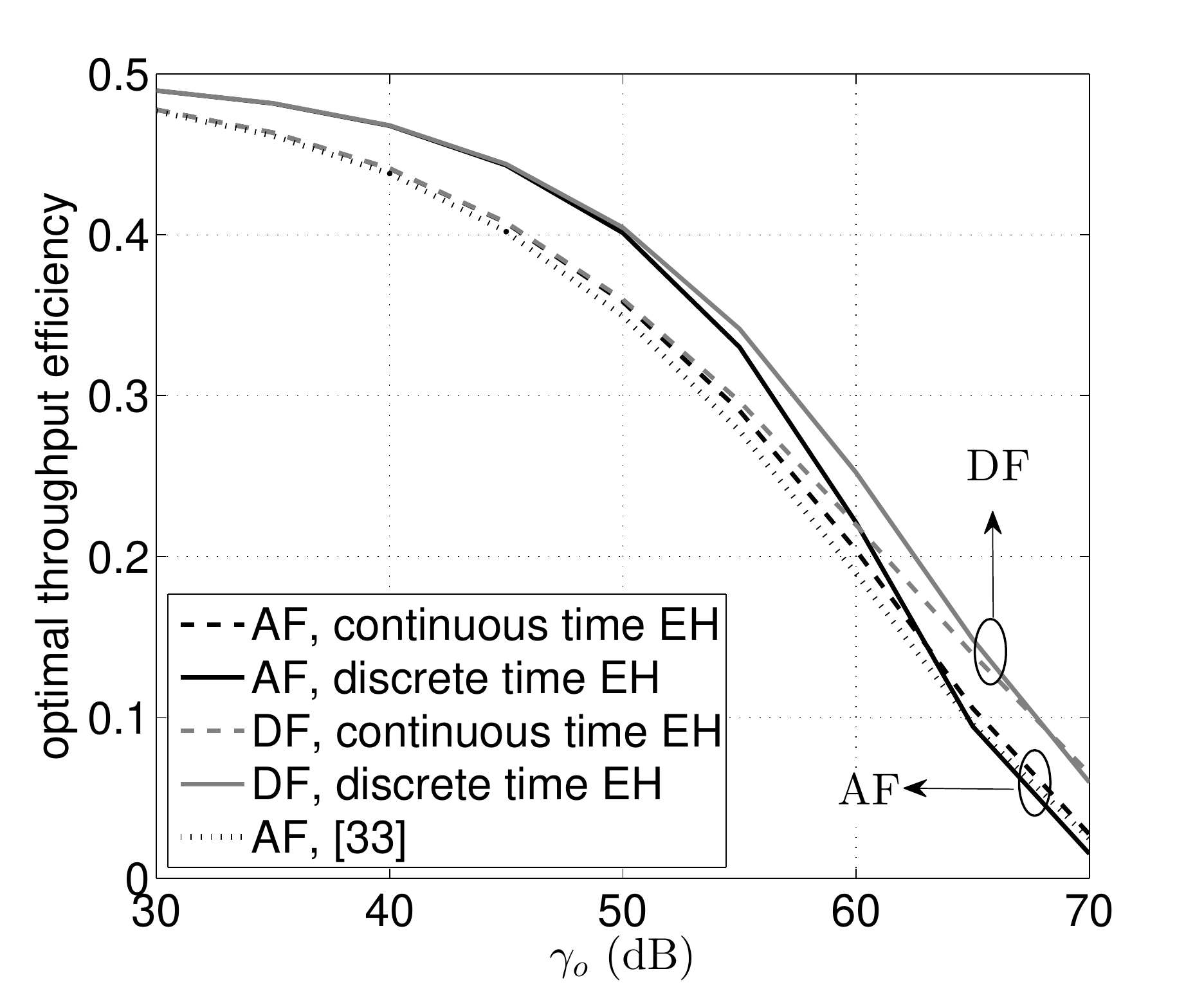}
  \caption{Optimal throughput efficiency versus detection threshold SNR, $\gamma_o$ for the proposed protocols with continuous time and discrete time EH and the fixed time-duration EH protocol in \cite{Nasir-13-A}.}
  \label{fig:Fig6}
\end{figure}

Fig. \ref{fig:Fig6} plots the optimal throughput for Protocols 1-4 and the  protocol in \cite{Nasir-13-A} as a function of SNR threshold $\gamma_o$. We can see from Fig. \ref{fig:Fig6} that the proposed continuous time EH outperforms fixed time-duration EH in \cite{Nasir-13-A} by approx. $0.2-0.5$ dB for a wide range of the threshold SNR ($\gamma_o \in (40,70)$ dBs). Similarly, the discrete time EH outperforms fixed time-duration EH in \cite{Nasir-13-A} by almost $3-7$ dB margin for SNR threshold $\gamma_o \in (30,60)$ dB. This performance improvement for the proposed protocols is due to the more efficient use of resources and can be intuitively explained as follows. With the fixed time-duration EH in \cite{Nasir-13-A}, the harvested energy in each block is not controllable and depends on the fading channel quality of the $\mathbb{S}$$\to$$\mathbb{R}$ link. When the $\mathbb{S}$$\to$$\mathbb{R}$ link is in deep fade, the harvested energy for the relay transmission will be very small which results in outage at the destination due to insufficient relaying power, $P_r$. When the $\mathbb{S}$$\to$$\mathbb{R}$ link is very strong, the harvested energy will be very large, which although guarantees a successful relay transmission but is a waste of energy. On the other hand, the proposed protocols allow relay energy accumulation and adapt the EH time to meet a preset relaying power which can be carefully chosen to give energy-and-throughput efficient performance.

Fig. \ref{fig:Fig6} shows that the optimal throughput efficiency of the proposed discrete time EH protocols is around $0.35-0.4$ at $\gamma_o$ = 50 dB. From the definition of the optimal throughput efficiency, this implies that on average $40\%$ of the block time is used for successful information transmission. This further implies that for discrete time EH protocols, if a certain block is used for EH, the probability that the next block will be used for IT is about $80\%$, on average. This result serves to demonstrate the feasibility of the proposed energy harvesting protocols.

It must be noted that the performance improvement of the proposed protocols comes at a marginal expense that the relay needs to alert the source and destination nodes for IT by sending a single bit, which is negligible overhead as compared to the amount of data being communicated. In addition, from an implementation complexity perspective, the proposed protocols are similar to the protocol in \cite{Nasir-13-A}. The proposed protocols vary EH time-duration to meet a fixed $P_r$, while the protocol in \cite{Nasir-13-A} varies $P_r$ to meet fixed EH time-duration for each block. Consequently, for the proposed protocols, while the relay has to monitor its available energy to decide when to start IT, the relay always transmits at the same power $P_r$, which eases the hardware design at the relay. On the other hand, the variable transmission power for the protocol in \cite{Nasir-13-A} increases hardware complexity and may require a large dynamic range of the power amplifier at the relay \cite{Ryhove-08-A}.
%

\section{Conclusions}\label{sec:conclusions}

In this paper, we have proposed TS-based protocols for EH and IT with continuous time and discrete time EH in AF and DF relaying. The proposed protocols adapt the EH time-duration at the relay node based on the available harvested energy and source-to-relay channel quality. We also derived the analytical expressions for the achievable throughput for the proposed protocols and verified them by comparisons with simulations. Many extensions are possible for the work presented in this paper. For example, the proposed protocols in this paper can be extended by adapting the EH time-duration opportunistically, i.e., the relay node can charge its battery to a level that is much more than what is required to transmit the source packet, depending on the channel conditions and then the extra harvested energy can be utilized opportunistically.

\appendices
\numberwithin{equation}{section}

\section{\vspace{-0pt}Proof of Theorem \ref{Th1} in \eqref{eq:tau_AFc_ana}}\label{app:A}
This appendix derives the analytical expression for the throughput $\tau$, in \eqref{eq:tau_AFc_ana}.

\noindent \emph{Step 1:} Using \eqref{eq:tau_AFc}, the throughput $\tau$ can be written as
\begin{align}\label{eq:tau_app1}
\tau &= \mathbb{E}_{h_i,g_i} \left\{ \frac{(1 - I_{o,i}) (1 - \alpha_i)}{2} \right\}.
\end{align}
Since, EH time, $\alpha_i$ in \eqref{eq:alpha_t_AFc}, is independent of $g_i$, \eqref{eq:tau_app1} can be written as
\begin{align}\label{eq:tau_app1_ext}
\tau &= \mathbb{E}_{h_i} \left\{  \frac{ \mathbb{E}_{g_i} \left\{1 - I_{o,i}\right\} (1 - \alpha_i)}{2} \right\}.
\end{align}

\noindent \emph{Step 2:} Let us first evaluate the inner expectation in \eqref{eq:tau_app1_ext}. By substituting \eqref{eq:gD_AF} into \eqref{eq:pout_AF}, $I_{o,i}$, is given by
\begin{align}\label{eq:pout_app1}
      I_{o,i} = \mathbbm{1} \left( \frac{P_s P_r |h_i|^2 |g_i|^2 }{P_r |g_i|^2 d_1^m \sigr + d_2^m \sigd ( P_s |h_i|^2  + d_1^m \sigr ) } < \gamma_o \right).
\end{align}
For notational convenience, we define $a \triangleq P_s d_2^m \sigd \gamma_o$, $b \triangleq d_1^{m} d_2^m \sigr \sigd \gamma_o$, $c \triangleq P_s P_r$, and $d \triangleq P_r d_1^m \sigr \gamma_o$. Using these variables in \eqref{eq:pout_app1} and simplifying, we get
\ifCLASSOPTIONpeerreview
\begin{align}\label{eq:pout_app2}
      I_{o,i} = \mathbbm{1} \left( (c |h_i|^2 - d) |g_i|^2 < (a|h_i|^2+b) \right)
       = \begin{cases}
                    \mathbbm{1} \left( |g_i|^2 < \frac{a|h_i|^2+b}{c |h_i|^2 - d} \right) , & |h_i|^2 > d/c \\
                    \mathbbm{1} \left( |g_i|^2 > \frac{a|h_i|^2+b}{c |h_i|^2 - d} \right) = 1 , & |h_i|^2 < d/c
       \end{cases}
\end{align}
\else
\begin{align}\label{eq:pout_app2}
      I_{o,i} &= \mathbbm{1} \left( (c |h_i|^2 - d) |g_i|^2 < (a|h_i|^2+b) \right) \notag \\
       &= \begin{cases}
                    \mathbbm{1} \left( |g_i|^2 < \frac{a|h_i|^2+b}{c |h_i|^2 - d} \right) , & |h_i|^2 > d/c \\
                    \mathbbm{1} \left( |g_i|^2 > \frac{a|h_i|^2+b}{c |h_i|^2 - d} \right) = 1 , & |h_i|^2 < d/c
       \end{cases}
\end{align}
\fi
As mentioned in Section \ref{sec:sys_mod}, $|g_i|^2$ is an exponential random variable with unit mean, i.e., $\mathbb{E}_{g_i} \left\{ \mathbbm{1} (\g < z) \right\} \triangleq \left(1 - e^{-z} \right) \mathbbm{1}(z)$, thus, $\mathbb{E}_{g_i} \left\{1 - I_{o,i}\right\}$ is given by
\ifCLASSOPTIONpeerreview
\begin{align}\label{eq:Eg_1pout}
      \mathbb{E}_{g_i} \left\{1 - I_{o,i}\right\} = \begin{cases}
                    e^{-\frac{a \h + b}{c \h - d }} , & |h_i|^2 > d/c \\
                    0 , & |h_i|^2 < d/c
       \end{cases} =  e^{-\frac{a \h + b}{c \h - d }} \mathbbm{1} \left( |h_i|^2 > d/c \right).
\end{align}
\else
\begin{align}\label{eq:Eg_1pout}
      \mathbb{E}_{g_i} \left\{1 - I_{o,i}\right\} &= \begin{cases}
                    e^{-\frac{a \h + b}{c \h - d }} , & |h_i|^2 > d/c \\
                    0 , & |h_i|^2 < d/c
       \end{cases} \notag  \\ &=  e^{-\frac{a \h + b}{c \h - d }} \mathbbm{1} \left( |h_i|^2 > d/c \right).
\end{align}
\fi

\noindent \emph{Step 3:} Next, we evaluate the outer expectation in \eqref{eq:tau_app1_ext} to get the throughput. Using \eqref{eq:alpha_t_AFc}, \eqref{eq:Eg_1pout}, and the fact that $\h$ is an exponential random variable with unit mean and the probability density function (PDF) $f_{h_i}(z) \triangleq  e^{-z}$, the throughput, $\tau$ in \eqref{eq:tau_app1_ext}, can be simplified to
\ifCLASSOPTIONpeerreview
\begin{align}\label{eq:tau_app1_simp1}
      \tau = \int_{d/c}^{\infty} \frac{1 - \alpha_i}{2} e^{- \frac{az+b}{cz-d} } f_{h_i}(z) dz = \int_{d/c}^{\infty} \frac{e^{ -\left( \frac{az+b}{cz-d} + z \right) }}{2}  dz - \frac{d_1^m P_r}{2 } \int_{d/c}^{\infty} \frac{ e^{ -\left( \frac{az+b}{cz-d} + z \right) }}{2 \eta P_s z + d_1^m P_r} dz.
\end{align}
\else
\begin{align}\label{eq:tau_app1_simp1}
      \tau &= \int_{d/c}^{\infty} \frac{1 - \alpha_i}{2} e^{- \frac{az+b}{cz-d} } f_{h_i}(z) dz \notag \\ &= \int_{d/c}^{\infty} \frac{e^{ -\left( \frac{az+b}{cz-d} + z \right) }}{2}  dz - \frac{d_1^m P_r}{2 } \int_{d/c}^{\infty} \frac{ e^{ -\left( \frac{az+b}{cz-d} + z \right) }}{2 \eta P_s z + d_1^m P_r} dz.
\end{align}
\fi
Defining a new integration variable $\bz \triangleq z- d/c$, the throughput $\tau$ can be written as
\ifCLASSOPTIONpeerreview
\begin{align} \label{eq:tau_app1_simp4}
      \tau = \frac{e^{- \frac{a+d}{c} }}{2 } \int_{0}^{\infty} e^{ -\left( \bz + \frac{ad+bc}{c^2 \bz} \right) }  d \bz - \frac{ e^{- \frac{a+d}{c} }}{2 } \int_{0}^{\infty} \frac{c d_1^m P_r e^{-\left( \bz + \frac{ad+bc}{c^2 \bz } \right)}}{2 c \eta P_s \bz + 2 \eta P_s d + c d_1^m P_r} d\bz  = \frac{e^{-  \frac{a+d}{c} }}{2} \left( u K_{1}(u) - c d_1^m P_r \nu \right),
\end{align}
\else
\begin{align} \label{eq:tau_app1_simp4}
      \tau &= \frac{e^{- \frac{a+d}{c} }}{2 } \int_{0}^{\infty} e^{ -\left( \bz + \frac{ad+bc}{c^2 \bz} \right) }   d \bz \notag \\ & \hspace{1.8cm} - \frac{ e^{- \frac{a+d}{c} }}{2 } \int_{0}^{\infty} \frac{c d_1^m P_r e^{-\left( \bz + \frac{ad+bc}{c^2 \bz } \right)}}{2 c \eta P_s \bz + 2 \eta P_s d + c d_1^m P_r} d\bz  \notag \\ &= \frac{e^{-  \frac{a+d}{c} }}{2} \left( u K_{1}(u) - c d_1^m P_r \nu \right),
\end{align}
\fi
where $u \triangleq \sqrt{\frac{4(ad+bc)}{c^2} }$, $\nu \triangleq \int_{x=0}^{\infty} \frac{e^{-\left( x + \frac{ad+bc}{c^2 x} \right)}}{2 c \eta P_s x + 2 \eta P_s d + c d_1^m P_r} dx$, and the last equality in \eqref{eq:tau_app1_simp4} is obtained by using the formula, $\int_{0}^{\infty} e^{-\frac{\beta}{4x} - \gamma x} dx = \sqrt{\frac{\beta}{\gamma}} K_{1}( \sqrt{\beta \gamma } )$ \cite[$\S$3.324.1]{Gradshteyn-80-B}. This completes the proof for Theorem \ref{Th1}.

\section{\vspace{-0pt}Proof of Lemma \ref{L1} in \eqref{eq:f_Eo_AFd}}\label{app:AB}

This appendix proves that the available harvested energy at the start of any EH-IT pattern, $E_o$, in Protocol 2 is exponentially distributed. The harvested energy available at the start of any EH-IT pattern is the same as the left over harvested energy in the battery at the end of the previous EH-IT pattern. Hence, in order to derive the PDF of $E_o$ we have to derive the PDF of the left over harvested energy at the end of any EH-IT pattern. We proceed as follows:

\noindent \emph{Step 1:} Consider the \textit{first} EH-IT pattern at the beginning of the transmission, consisting of certain number of EH block(s) prior to an IT block. For this first EH-IT pattern, there will no (prior) accumulated energy. During each EH block, energy will be accumulated. Since $|h_i|^2$ is exponentially distributed with the PDF $f_{\h}(z) \triangleq e^{-z}$, the harvested energy per block interval, $E_i^{0 \to T} = \frac{\eta P_s |h_i|^2}{d_1^m} T$ in \eqref{eq:EH1_AFd}, is exponentially distributed with parameter $\frac{1}{\rho}$, i.e., $f_{E_i^{0 \to T}}(\epsilon) =  \frac{1}{\rho} e^{- \frac{\epsilon}{\rho} }$. The left over harvested energy at the end of the \textit{first} EH-IT pattern will be the extra harvested energy, above $\frac{P_r T}{2}$ level, in an EH block preceding an IT block because the IT block will consume a fixed $\frac{P_r T}{2}$ level of energy. Following the \emph{memoryless property} of the exponential distribution \cite[Ch. 2]{Gallager-96-B}, the extra harvested energy, over $\frac{P_r T}{2}$ level, in an EH block preceding an IT block is exponentially distributed with parameter $\frac{1}{\rho}$. Consequently, the left over harvested energy at the end of the \textit{first} EH-IT pattern is exponentially distributed with parameter $\frac{1}{\rho}$.

\noindent \emph{Step 2:} Now consider the \textit{second} EH-IT pattern in the transmission. From Step 1, we know that the left over harvested energy at the end of the first EH-IT pattern is exponentially distributed. Hence, for the \textit{second} EH-IT pattern there is available harvested energy at the start of the pattern and it is exponentially distributed. We show that given this available harvested energy at the start of the pattern which is exponentially distributed, the left over harvested energy at the end of the \textit{second} EH-IT pattern is also exponentially distributed with parameter $\frac{1}{\rho}$. To do this we need to consider the possible cases of the \textit{second} EH-IT pattern. The \textit{second} EH-IT pattern may consist of either EH block(s) followed by an IT block or a single IT block (see discussion in Section~\ref{sec:XXX}). These two cases are considered below:
    \begin{enumerate}
    \item If the \textit{second} EH-IT pattern contains EH block(s) followed by an IT block, we can use the same argument as in Step 1 above to prove that the left over harvested energy at the end of the \textit{second} EH-IT pattern is exponentially distributed with parameter $\frac{1}{\rho}$.

     \item If the \textit{second} EH-IT pattern consists of a single IT block, this implies that the initial energy $E_o$ for this \textit{second} EH-IT pattern is already greater than the threshold level $\frac{P_r T}{2}$. Now, in order to prove that the left over harvested energy in the battery at the end of the single IT block is exponentially distributed, we have to prove that PDF of $E_o - \frac{P_r T}{2}$, given $E_o > \frac{P_r T}{2}$ is exponential. In order to find $p \left(E_o - \frac{P_r T}{2} \big| E_o > \frac{P_r T}{2}\right)$, we use the fact that
    \ifCLASSOPTIONpeerreview
    \begin{align}\label{eq:CDF_AB}
     p \left(E_o - \frac{P_r T}{2} < z \bigg|E_o > \frac{P_r T}{2}\right) &=  p \left(E_o  < z + \frac{P_r T}{2} \bigg|E_o > \frac{P_r T}{2}\right) \notag \\ &= \begin{cases}
       0 , & z < 0 \\
       \frac{e^{-\frac{P_r T}{2 \rho}} - e^{-\frac{1}{\rho}\left( z + \frac{P_r T}{2} \right)} }{e^{-\frac{P_r T}{2 \rho}}}, & z> 0
     \end{cases}
    \end{align}
    \else
    \begin{align}\label{eq:CDF_AB}
     p \left(E_o - \frac{P_r T}{2} < z \bigg|E_o > \frac{P_r T}{2}\right) & \notag \\  & \hspace{-3cm} =  p \left(E_o  < z + \frac{P_r T}{2} \bigg|E_o > \frac{P_r T}{2}\right) \notag \\ & \hspace{-3cm} = \begin{cases}
       0 , & z < 0 \\
       \frac{e^{-\frac{P_r T}{2 \rho}} - e^{-\frac{1}{\rho}\left( z + \frac{P_r T}{2} \right)} }{e^{-\frac{P_r T}{2 \rho}}}, & z> 0
     \end{cases}
    \end{align}
    \fi
    where $p(\cdot)$ denotes probability. Using \eqref{eq:CDF_AB}, the PDF, $p \left(E_o - \frac{P_r T}{2} \big| E_o > \frac{P_r T}{2}\right)$, can be derived as
     \ifCLASSOPTIONpeerreview
    \begin{align}\label{eq:PDF_AB}
     p \left(E_o - \frac{P_r T}{2} \big| E_o > \frac{P_r T}{2}\right) &=  \frac{\partial}{\partial z} p \left(E_o - \frac{P_r T}{2} < z \bigg|E_o > \frac{P_r T}{2}\right) \notag \\ &= \begin{cases}
       0 , & z < 0 \\
       \frac{1}{\rho} e^{-\frac{z}{\rho}}, & z> 0
     \end{cases}
    \end{align}
    \else
    \begin{align}\label{eq:PDF_AB}
     p \left(E_o - \frac{P_r T}{2} \big| E_o > \frac{P_r T}{2}\right) & \notag \\  & \hspace{-3cm} =  \frac{\partial}{\partial z} p \left(E_o - \frac{P_r T}{2} < z \bigg|E_o > \frac{P_r T}{2}\right) \notag \\ & \hspace{-3cm} = \begin{cases}
       0 , & z < 0 \\
       \frac{1}{\rho} e^{-\frac{z}{\rho}}, & z> 0
     \end{cases}
    \end{align}
    \fi
    From \eqref{eq:PDF_AB}, we can conclude that if the \textit{second} EH-IT pattern consists of a single IT block, the PDF of the left over harvested energy at the end of the IT block is exponentially distributed with parameter $\frac{1}{\rho}$. Combining the results for the two cases above, we can conclude that in general, the left over harvested energy at the end of the \textit{second} EH-IT pattern is exponentially distributed with parameter $\frac{1}{\rho}$.
    \end{enumerate}

\noindent \emph{Step 3:} We can generalize from Steps 1 and 2 that the left over harvested energy at the end of \textit{any} EH-IT pattern, in a sequence of EH-IT patterns, is exponentially distributed. Consequently, the harvested energy available at the start of any EH-IT pattern, $E_o$, is an exponentially distributed with the PDF, $f_{E_o}(\epsilon) =  \frac{1}{\rho } e^{- \frac{\epsilon}{\rho } }$, where $\rho \triangleq \frac{\eta P_s T}{d_1^m}$. This completes the proof of Lemma \ref{L1}.

\ifCLASSOPTIONpeerreview

\else
 \begin{figure*}[!b]
 \hrulefill
 \vspace{-0.2cm}
\normalsize
\setcounter{MYtempeqncnt}{0}
\setcounter{equation}{0}
\begin{align}\label{eq:alpha_t_DFc_approx}
      \tilde{\alpha}_i \approx \begin{cases}
                    1 , & \h < \ba \\
                    \frac{d_1^m P_r}{2\eta P_s | h_i|^2  + d_1^m P_r }, & \h \ge \ba \;,\; |h_{i-1}|^2 \ge \ba \\
                    \frac{d_1^m P_r - 2 \eta P_s \sum_{k=1}^{n} |h_{i-k}|^2}{2\eta P_s | h_i|^2  + d_1^m P_r }, &  \h \ge \ba \;,\; |h_{i-k}|^2 < \ba \;\forall \; k = 1,\hdots,n  \;,\;  |h_{i-(n+1)}|^2 \ge \ba
       \end{cases} \tag{D.1}
\end{align}
\setcounter{equation}{\value{MYtempeqncnt}}
\vspace*{-.2cm}
\end{figure*}
\fi

\ifCLASSOPTIONpeerreview

\else
 \begin{figure*}[!b]
 \hrulefill
\normalsize
\setcounter{MYtempeqncnt}{0}
\setcounter{equation}{0}
\begin{align}\label{eq:tau_t_DFc_approx}
      \tilde{\tau}_i  \approx   \begin{cases}
                    0 , &  \h < \ba \;||\; \g < \bb \\
                    \frac{\chi_{0,i}}{2}, &  \h \ge \ba \;,\; \g \ge \bb \;,\; |h(t-T)|^2 \ge \ba \\
                    \frac{\chi_{n,i}}{2} , &   \h \ge \ba \;,\; \g \ge \bb \;,\; |h_{i-k}|^2 < \ba \; \forall \; k = 1,\hdots,n  \;,\;  |h_{i-(n+1)}|^2 \ge \ba
       \end{cases} \tag{D.2}
\end{align}
\setcounter{equation}{\value{MYtempeqncnt}}
\vspace*{-.2cm}
\end{figure*}

\fi

\section{\vspace{-0pt}Proof of Theorem \ref{Th2} in \eqref{eq:tau_AFd_ana}}\label{app:B}
This appendix derives the analytical expression for the throughput $\tau$, in \eqref{eq:tau_AFd_ana}.

\noindent \emph{Step 1:} First we have to simplify $\tau = \mathbb{E}_{g_i,\mathbf{h}} \left\{ \frac{(1 - I_{o,i}) R (1 - \alpha_i)}{2} \right\}$, where $I_{o,i}$ and $\alpha_i$ are given in \eqref{eq:pout_AF} and \eqref{eq:alpha_t_AFd}, respectively. In Protocol 2, block outage indicator, $I_{o,i}$, is independent of the EH time, $\alpha_i$. This is because $I_{o,i}$ depends on the fading channels, $h_i$ and $g_i$, of the current block. However, $\alpha_i$ depends on the accumulated harvested energy at the start of the block, $E_i(0)$ (see \eqref{eq:alpha_t_AFd}), which in turn depends on the $\mathbb{S}-\mathbb{R}$ fading channels of the previous blocks, i.e., $\mathbf{h} \setminus h_i = \{h_{i-1},h_{i-2},\hdots\}$. Thus, throughput, $\tau$, is given by
\begin{align}\label{eq:tau_app_b1}
\tau = \frac{1}{2} \hspace{0.2cm} \mathbb{E}_{h_i,g_i} \left\{1 - I_{o,i}\right\} \hspace{0.2cm} \mathbb{E}_{\mathbf{h} \setminus h_i} \{1 - \alpha_i\}.
\end{align}

\noindent \emph{Step 2:} We have two expectations to evaluate. Using \eqref{eq:Eg_1pout} the first expectation in \eqref{eq:tau_app_b1}, $\mathbb{E}_{h_i,g_i} \left\{1 - I_{o,i}\right\}$ is given by
\ifCLASSOPTIONpeerreview
\begin{align}\label{eq:Egh_1pout}
      \mathbb{E}_{h_i,g_i} \left\{1 - I_{o,i}\right\} = \frac{1}{\lh} \int_{d/c}^{\infty} e^{- \left(\frac{az+b}{(cz-d)\lgg} + \frac{z}{\lh} \right)}  dz = e^{-\left( \frac{a}{c \lgg} + \frac{d}{c \lh} \right)} u K_{1} \left(  u \right).
\end{align}
\else
\begin{align}\label{eq:Egh_1pout}
      \mathbb{E}_{h_i,g_i} \left\{1 - I_{o,i}\right\} &= \frac{1}{\lh} \int_{d/c}^{\infty} e^{- \left(\frac{az+b}{(cz-d)\lgg} + \frac{z}{\lh} \right)}  dz \notag \\ &= e^{-\left( \frac{a}{c \lgg} + \frac{d}{c \lh} \right)} u K_{1} \left(  u \right).
\end{align}
\fi

\noindent \emph{Step 3:} The second expectation, $\mathbb{E}_{\mathbf{h} \setminus h_i} \{1 - \alpha_i\}$ is the expected value that any block is an IT block. From the general EH-IT pattern (a), given in Fig. \ref{fig:AFd}, probability of any block being IT block can be written as
\begin{align}\label{eq:E_1alpha}
      \mathbb{E}_{\mathbf{h} \setminus h_i} \{1 - \alpha_i\} = \frac{1}{\mathbb{E}_{\bar{X}} \{X+1\}},
\end{align}
where $\bar{X} \triangleq X-1$. Using $f_{E_o}(\epsilon)$ and $p_{\bar{X}|E_o}(\bar{x}|E_o)$ from \eqref{eq:f_Eo_AFd} and \eqref{eq:pX_AFd}, respectively, $\mathbb{E}_{\bar{X}} \{X\}$ is given by
\ifCLASSOPTIONpeerreview
\begin{subequations}\label{eq:E_X}
\begin{align}
      \mathbb{E}_{\bar{X}} \{X\} &= \mathbb{E}_{E_o}  \left\{ \mathbb{E}_{\bar{X}|E_o} \{ X|E_o \}  \right\} = \int_{\epsilon = 0}^{P_r T/2}  \mathbb{E}_{\bar{X}|E_o} \{ \bar{X}+1|E_o \}  \; f_{E_o}(\epsilon) d \epsilon + \int_{\epsilon = P_r T/2}^{\infty}  0 d \epsilon \label{eq:E_X1} \\
      &= \int_{\epsilon = 0}^{P_r T/2}  \left(\frac{1}{\rho } \left( \frac{P_r T}{2} - \epsilon \right) + 1\right) f_{E_o}(\epsilon) d \epsilon = \frac{P_r d_1^m}{2 \eta P_s \lh} \label{eq:E_X2}.
\end{align}
\end{subequations}
\else
\begin{subequations}\label{eq:E_X}
\begin{align}
      \mathbb{E}_{\bar{X}} \{X\} &= \mathbb{E}_{E_o}  \left\{ \mathbb{E}_{\bar{X}|E_o} \{ X|E_o \}  \right\} \notag \\ &= \int_{\epsilon = 0}^{P_r T/2}  \mathbb{E}_{\bar{X}|E_o} \{ \bar{X}+1|E_o \}  \; f_{E_o}(\epsilon) d \epsilon + \int_{\epsilon = P_r T/2}^{\infty}  0 d \epsilon \label{eq:E_X1} \\
      &= \int_{\epsilon = 0}^{P_r T/2}  \left(\frac{1}{\rho } \left( \frac{P_r T}{2} - \epsilon \right) + 1\right) f_{E_o}(\epsilon) d \epsilon \notag \\ &= \frac{P_r d_1^m}{2 \eta P_s \lh} \label{eq:E_X2}.
\end{align}
\end{subequations}
\fi
%

\noindent \emph{Step 4:} Substituting \eqref{eq:Egh_1pout}, \eqref{eq:E_1alpha}, and \eqref{eq:E_X2} into \eqref{eq:tau_app_b1}, we can obtain the analytical throughput expression as given in \eqref{eq:tau_AFd_ana}. This completes the proof for Theorem \ref{Th2}.

\ifCLASSOPTIONpeerreview

\else
 \begin{figure*}[!b]
  \hrulefill
 \vspace{-0.0cm}
\normalsize
\setcounter{MYtempeqncnt}{0}
\setcounter{equation}{0}
\begin{align}\label{eq:integ_DFc}
 \underbrace{\int_{0}^{\ba} \cdots \int_{0}^{\ba} \hspace{-0.2cm} \int_{\ba}^{\infty}}_{n+1 \text{integrals}}  \chi_{n,i}  e^{( h_i+\cdots+h_n ) } d h_i \hdots dh_{i-n} &= \frac{1}{2 \eta P_s} \left( e^{- \ba}  ( 1 - e^{- \ba})^{n-1} ( 1 - e^{- q}) - e^{\left( \frac{d_1^m P_r - 2n \eta P_s ( (n-1)\ba + q)}{2 \eta P_s } \right)}  \right. \notag \\ & \hspace{-3.5cm} \times \left. \left(  e^{ \ba} - 1 \right)^{n-1} \left(2 n \eta P_sq - d_1^m P_r + (d_1^m P_r - 2n \eta P_s ) e^{q} \right) E_1\left( \frac{d_1^m P_r + 2\eta P_s \ba}{2 \eta P_s } \right) \right),  \tag{D.4}
\end{align}
\setcounter{equation}{\value{MYtempeqncnt}}
\vspace*{-.4cm}
\end{figure*}
\fi

\section{\vspace{-0pt}Proof of Theorem \ref{Th3} in \eqref{eq:tau_DFc_ana}}\label{app:C}
This appendix derives the analytical expression for the lower bound throughput $\tau$, in \eqref{eq:tau_DFc_ana} using the assumption $E_o = 0$. We use the notation $(\tilde{\cdot})$ to denote a quantity which is calculated using this assumption.

\noindent \emph{Step 1:} The assumption $E_o = 0$ implies that $E_i(0)$ in \eqref{eq:alpha_t_DFc} is equal to $0$ if $n=0$, else it is equal to $ \frac{\eta P_s \sum_{k=1}^{n} |h_{i-k}|^2 T }{ d_1^m}$ if $n\neq 0$, where $n$ denotes the number of successive EH blocks due to relay outage. Thus, \eqref{eq:alpha_t_DFc} can be expressed by
\ifCLASSOPTIONpeerreview
\begin{align}\label{eq:alpha_t_DFc_approx}
      \tilde{\alpha}_i = \begin{cases}
                    1 , & \h < \ba \\
                    \frac{d_1^m P_r}{2\eta P_s | h_i|^2  + d_1^m P_r }, & \h \ge \ba \;,\; |h_{i-1}|^2 \ge \ba \\
                    \frac{d_1^m P_r - 2 \eta P_s \sum_{k=1}^{n} |h_{i-k}|^2}{2\eta P_s | h_i|^2  + d_1^m P_r }, &  \h \ge \ba \;,\; |h_{i-k}|^2 < \ba \;\forall \; k = 1,\hdots,n  \;,\;  |h_{i-(n+1)}|^2 \ge \ba
       \end{cases}
\end{align}
\else
\addtocounter{equation}{1}
\eqref{eq:alpha_t_DFc_approx} at the bottom of the page,
\fi

\noindent \emph{Step 2:} Using \eqref{eq:tau_DF}, \eqref{eq:alpha_t_DFc_approx}, and $E_o = 0$, the general block throughput can be expressed as
\ifCLASSOPTIONpeerreview
\begin{align}\label{eq:tau_t_DFc_approx}
      \tilde{\tau}_i  =   \begin{cases}
                    0 , &  \h < \ba \;||\; \g < \bb \\
                    \frac{\chi_{0,i}}{2}, &  \h \ge \ba \;,\; \g \ge \bb \;,\; |h(t-T)|^2 \ge \ba \\
                    \frac{\chi_{n,i}}{2} , &   \h \ge \ba \;,\; \g \ge \bb \;,\; |h_{i-k}|^2 < \ba \; \forall \; k = 1,\hdots,n  \;,\;  |h_{i-(n+1)}|^2 \ge \ba
       \end{cases}
\end{align}
\else
\addtocounter{equation}{1}
\eqref{eq:tau_t_DFc_approx} at the bottom of the page,
\fi
where $||$ is the logical OR operator, $\chi_{0,i} \triangleq 1 - \frac{d_1^m P_r}{2\eta P_s | h_i|^2  + d_1^m P_r }$  and $\chi_{n,i} \triangleq \min \left( 1 , 1 - \frac{d_1^m P_r - 2 \eta P_s \sum_{k=1}^{n} |h_{i-k}|^2}{2\eta P_s | h_i|^2  + d_1^m P_r } \right)$ for $n > 0$.

\noindent \emph{Step 3:}  Let us define $\mathbf{h}_n \triangleq \{ h_i, h_{i-1},\hdots, h_{i-n} \}$. From \eqref{eq:tau_t_DFc_approx}, $\tilde{\tau}_i$ depends on $\mathbf{h}_n$ and $g_i$. Given that all the absolute-squared channel gains are exponentially distributed, the throughput, $\tilde{\tau}$ can be determined as follows.
\ifCLASSOPTIONpeerreview
\begin{align}\label{eq:tau_der_DFc}
\tilde{\tau} &= \mathbb{E}_{\mathbf{h}_n,g(t)} \{ \tilde{\tau}_i \} = \mathbb{E}_{\mathbf{h}_n} \left\{ \tilde{\tau}_i|_{g_i < \bb} \;\; p(g_i < \bb) + \tilde{\tau}_i|_{g_i \ge \bb} \;\; p(g_i \ge \bb) \right\} =  e^{-\bb } \mathbb{E}_{\mathbf{h}_n} \left\{ \tilde{\tau}_i|_{g_i < \bb} \right\} \notag \\
     &=  e^{-\bb } \mathbb{E}_{\mathbf{h}_n \setminus h_i} \left\{ \int_{\ba}^{\infty} \tilde{\tau}_i|_{g_i \ge \bb,h_i \ge \ba} \;\;  e^{h_i } d h_i \right\} \notag  \\
     &=  e^{-\bb } \mathbb{E}_{\mathbf{h}_n \setminus h_i,h_{i-1}} \left\{ \int_{\ba}^{\infty} \tilde{\tau}_i|_{g_i \ge \bb,h_i \ge \ba,h_{i-1} \ge \ba} \; e^{h_i } d h_i \; p(h_{i-1} \ge \ba) + \int_{0}^{\ba} \int_{\ba}^{\infty} \tilde{\tau}_i|_{g_i \ge \bb,h_i \ge \ba,h_{i-1} < \ba} e^{- (h_i+h_{i-1}) } dh_i dh_{i-1} \right\} \notag   \\
     &=  e^{-\bb } \Bigg( e^{-\ba }  \int_{\ba}^{\infty} \frac{\chi_{0,i}}{2}   e^{h_i } d h_i  +  \mathbb{E}_{\mathbf{h}_n \setminus h_i,h_{i-1},h_{i-2}} \Bigg\{ \int_{0}^{\ba} \int_{\ba}^{\infty} \underbrace{\tilde{\tau}_i|_{g_i \ge \bb,h_i \ge \ba,h_{i-1} < \ba,h_{i-2} \ge \ba}}_{= \frac{\chi_{1,i}}{2}}   e^{- (h_i+h_{i-1}) } dh_i dh_{i-1} \;  p(h_{i-2} \ge \ba)  \notag  \\ & \hspace{6.2cm} + \int_{0}^{\ba} \int_{0}^{\ba} \int_{\ba}^{\infty} \tilde{\tau}_i|_{g_i \ge \bb,h_i \ge \ba,h_{i-1} < \ba, h_{i-2} < \ba}   e^{- (h_i+h_{i-1}+h_{i-2}) } dh_i dh_{i-1} dh_{i-2} \Bigg\} \Bigg) \notag \\
     &= \frac{e^{-(\ba+\bb)}}{2} \left(  \int_{\ba}^{\infty}  \chi_{0,i}  e^{h_i } d h_i + \int_{0}^{\ba} \hspace{-0.2cm} \int_{\ba}^{\infty} \chi_{1,i}  e^{- (h_i+h_{i-1}) } d h_i dh_{i-1} +  \int_{0}^{\ba} \hspace{-0.2cm} \int_{0}^{\ba} \hspace{-0.2cm} \int_{\ba}^{\infty} \chi_{2,i}  e^{- (h_i+h_{i-1}+h_{i-2}) } d h_i dh_{i-1} dh_{i-2} + \hspace{-0.1cm} \cdots \hspace{-0.1cm} \right),
\end{align}
\else
\begin{align}\label{eq:tau_der_DFc}
\tilde{\tau} &= \mathbb{E}_{\mathbf{h}_n,g_i} \{ \tilde{\tau}_i \} \notag \\  &= \mathbb{E}_{\mathbf{h}_n} \left\{ \tilde{\tau}_i|_{g_i < \bb} \;\; p(g_i < \bb) + \tilde{\tau}_i|_{g_i \ge \bb} \;\; p(g_i \ge \bb) \right\} \notag \\  &= e^{-\bb } \mathbb{E}_{\mathbf{h}_n} \left\{ \tilde{\tau}_i|_{g_i \ge \bb} \right\} \notag \\
     &=  e^{-\bb } \mathbb{E}_{\mathbf{h}_n \setminus h_i} \left\{ \int_{\ba}^{\infty} \tilde{\tau}_i|_{g_i \ge \bb,h_i \ge \ba} \;\;  e^{h_i } d h_i \right\} \notag \\
     &=  e^{-\bb } \mathbb{E}_{\mathbf{h}_n \setminus h_i,h_{i-1}} \hspace{-0.12cm} \left\{ \int_{\ba}^{\infty} \tilde{\tau}_i|_{g_i \ge \bb,h_i \ge \ba,h_{i-1} \ge \ba} \; e^{h_i } d h_i  p(h_{i-1} \ge \ba) \right. \notag \\ & \hspace{1.2cm} \left. + \int_{0}^{\ba} \int_{\ba}^{\infty} \tilde{\tau}_i|_{g_i \ge \bb,h_i \ge \ba,h_{i-1} < \ba} e^{- (h_i+h_{i-1}) } dh_i dh_{i-1} \right\} \notag  \displaybreak \\
     &=  e^{-\bb } \Bigg( e^{-\ba }  \int_{\ba}^{\infty} \frac{\chi_{0,i}}{2}   e^{h_i } d h_i  +  \mathbb{E}_{\mathbf{h}_n \setminus h_i,h_{i-1},h_{i-2}} \Bigg\{  \int_{0}^{\ba} \int_{\ba}^{\infty} \notag \\ & \hspace{0.8cm} \underbrace{\tilde{\tau}_i|_{g_i \ge \bb,h_i \ge \ba,h_{i-1} < \ba,h_{i-2} \ge \ba}}_{= \frac{\chi_{1,i}}{2}}   e^{- (h_i+h_{i-1}) } dh_i dh_{i-1} \notag \\ & \hspace{0.8cm} \times  p(h_{i-2} \ge \ba)   + \int_{0}^{\ba} \int_{0}^{\ba} \int_{\ba}^{\infty} \tilde{\tau}_i|_{g_i \ge \bb,h_i \ge \ba,h_{i-1} < \ba, h_{i-2} < \ba}  \notag \\   & \hspace{2.7cm} \times  e^{- (h_i+h_{i-1}+h_{i-2}) } dh_i dh_{i-1} dh_{i-2} \Bigg\} \Bigg) \notag \\ &= \frac{e^{-(\ba+\bb)}}{2} \left(  \int_{\ba}^{\infty}  \chi_{0,i}  e^{h_i } d h_i + \int_{0}^{\ba} \hspace{-0.2cm} \int_{\ba}^{\infty} \chi_{1,i}  e^{- (h_i+h_{i-1}) } d h_i dh_{i-1} \right. \notag \\ & \left. + \hspace{0.2cm}  \int_{0}^{\ba}  \int_{0}^{\ba}  \int_{\ba}^{\infty} \chi_{2,i}  e^{- (h_i+h_{i-1}+h_{i-2}) } d h_i dh_{i-1} dh_{i-2} + \cdots  \right),
\end{align}
\fi
where
\ifCLASSOPTIONpeerreview
\begin{align}\label{eq:integ_DFc}
 \underbrace{\int_{0}^{\ba} \cdots \int_{0}^{\ba} \hspace{-0.2cm} \int_{\ba}^{\infty}}_{n+1 \text{integrals}}  \chi_{n,i}  e^{( h_i+\cdots+h_n ) } d h_i \hdots dh_{i-n} &= \frac{1}{2 \eta P_s} \left( e^{- \ba}  ( 1 - e^{- \ba})^{n-1} ( 1 - e^{- q}) - e^{\left( \frac{d_1^m P_r - 2n \eta P_s ( (n-1)\ba + q)}{2 \eta P_s } \right)}  \right. \notag \\ & \hspace{-3.5cm} \times \left. \left(  e^{ \ba} - 1 \right)^{n-1} \left(2 n \eta P_sq - d_1^m P_r + (d_1^m P_r - 2n \eta P_s ) e^{q} \right) E_1\left( \frac{d_1^m P_r + 2\eta P_s \ba}{2 \eta P_s } \right) \right),
\end{align}
\else
$\underbrace{\int_{0}^{\ba} \cdots \int_{0}^{\ba} \hspace{-0.2cm} \int_{\ba}^{\infty}}_{n+1 \text{integrals}}  \chi_{n,i}  e^{( h_i+\cdots+h_n ) } d h_i \hdots dh_{i-n}$ is given at the bottom of the page in \eqref{eq:integ_DFc},
\fi
and \eqref{eq:tau_der_DFc} is obtained by taking the expectation over all the elements of the set $\{ h_i,\hdots,h_{i-n},g_i \}$, one by one.

\noindent \emph{Step 4:} Substituting \eqref{eq:integ_DFc} into \eqref{eq:tau_der_DFc} results in the lower bound given in \eqref{eq:tau_DFc_ana}. Note that $\tau \ge \tilde{\tau}$ because of the assumption $E_o = 0$ used in the analysis. This completes the proof for Theorem \ref{Th3}.

\section{\vspace{-0pt}Proof of Theorem \ref{Th4} in \eqref{eq:tau_DFd_ana}}\label{app:D}
This appendix derives the analytical expression for the lower bound throughput $\tau$, in \eqref{eq:tau_DFd_ana}.

\noindent \emph{Step 1:} In the block throughput expression, $\tau_i = \frac{(1 - I_{o,i}) (1 - \alpha_i)}{2}$, the block outage indicator, $I_{o,i}$, is independent of the EH time $\alpha_i$. This is because $I_{o,i}$, given in \eqref{eq:pout_DF}, depends on the fading channel $g_i$. However, $\alpha_i$ in \eqref{eq:alpha_t_DFd}, depends on the $\mathbb{S}-\mathbb{R}$ fading channel, $h_i$. Also, $\alpha_i$ depends on the accumulated harvested energy $E_i(0)$ at the start of the block (see \eqref{eq:alpha_t_DFd}), which in turn depends on the $\mathbb{S}-\mathbb{R}$ channel of the previous blocks, i.e., $\{h_{i-1},h_{i-2},\hdots\}$. Thus, analytical throughput, $\tau$, is given by
\begin{align}\label{eq:tau_app_d1}
\tau &= \frac{1}{2} \hspace{0.2cm} \mathbb{E}_{g_i} \left\{1 - I_{o,i}\right\} \hspace{0.2cm} \mathbb{E}_{\mathbf{h}} \{1 - \alpha_i\}.
\end{align}

\noindent \emph{Step 2:} Using \eqref{eq:pout_DF}, the first expectation in \eqref{eq:tau_app_d1}, $\mathbb{E}_{\g} \left\{1 - I_{o,i}\right\}$ is given by
\begin{align}\label{eq:E_g_pdt}
\mathbb{E}_{g_i} \left\{1 - I_{o,i}\right\} = \mathbb{E}_{g_i} \left\{ \mathbbm{1} (\g > \bb) \right\} = e^{-\bb}
\end{align}

\noindent \emph{Step 3:} The second expectation in \eqref{eq:tau_app_d1}, $\mathbb{E}_{\mathbf{h}} \{1 - \alpha_i\}$ is the expected value that any block is an IT block. From general EH-IT pattern (a), given in Fig. \ref{fig:DFd}, probability of any block being an IT block is given by
\begin{align}\label{eq:E_1alpha_DFd}
      \mathbb{E}_{\mathbf{h}} \{1 - \alpha_i\} = \frac{1}{\mathbb{E}_{\bar{X},Y} \{X+Y+1\}},
\end{align}
Since the $\mathbb{S}-\mathbb{R}$ channel is independent for different blocks, $X$ and $Y$ are independent random variables. Thus, the denominator in \eqref{eq:E_1alpha_DFd} can be written as $\mathbb{E}_{\bar{X}}\{X\} +  \mathbb{E}_{Y}\{Y\}+1$, where $\mathbb{E}_{\bar{X}}\{X\}$ is given by
\ifCLASSOPTIONpeerreview
\begin{align}\label{eq:E_X_DF1}
      \mathbb{E}_{X} \{X\} = \mathbb{E}_{E_o}  \left\{ \mathbb{E}_{\bar{X}|E_o} \{ X|E_o \}  \right\} =   \int_{\epsilon = 0}^{P_r T/2} \mathbb{E}_{\bar{X}|E_o} \{ \bar{X}+1|E_o \}  f_{E_o}(\epsilon) d \epsilon,
\end{align}
\else
\begin{align}\label{eq:E_X_DF1}
      \mathbb{E}_{X} \{X\} &= \mathbb{E}_{E_o}  \left\{ \mathbb{E}_{\bar{X}|E_o} \{ X|E_o \}  \right\}  \notag \\  &=  \int_{\epsilon = 0}^{P_r T/2} \mathbb{E}_{\bar{X}|E_o} \{ \bar{X}+1|E_o \}  f_{E_o}(\epsilon) d \epsilon,
\end{align}
\fi
where $\mathbb{E}_{\bar{X}|E_o} \{ \bar{X}|E_o \} = \frac{1}{\rho } \left( \frac{P_r T}{2} - \epsilon \right) $ is defined below \eqref{eq:pX_AFd}.

Neglecting the harvested energy from $Y$ successive EH blocks due to relay outage, the pdf of $E_o$ becomes the same as that given in Lemma 1. Hence, \eqref{eq:E_X_DF1} simplifies to
\ifCLASSOPTIONpeerreview
\begin{align}\label{eq:E_X_DF2}
      \mathbb{E}_{X} \{X\} \le   \int_{\epsilon = 0}^{P_r T/2} \left(\frac{1}{\rho } \left( \frac{P_r T}{2} - \epsilon \right) + 1\right) \frac{1}{\rho } e^{- \frac{\epsilon}{\rho } } d \epsilon = \frac{P_r d_1^m}{2 \eta P_s \lh},
\end{align}
\else
\begin{align}\label{eq:E_X_DF2}
      \mathbb{E}_{X} \{X\} \le  \int_{\epsilon = 0}^{P_r T/2} \left(\frac{1}{\rho } \left( \frac{P_r T}{2} - \epsilon \right) + 1\right) \frac{1}{\rho } e^{- \frac{\epsilon}{\rho } } d \epsilon = \frac{P_r d_1^m}{2 \eta P_s \lh},
\end{align}
\fi
where the inequality sign in \eqref{eq:E_X_DF2} appears because for analysis we neglected the harvested energy from $Y$ successive EH blocks due to relay outage. Next, using the PMF of $Y$, $p_{Y}(y)$ in \eqref{eq:PMF_Y}, $\mathbb{E}_{Y}\{Y\}$ can be calculated as
\begin{align}\label{eq:E_Y}
\mathbb{E}_{Y}\{Y\} = \sum_{y=0}^{\infty} y (1 - p_{o,r}) p_{o,r}^y = \frac{p_{o,r}}{1 - p_{o,r}},
\end{align}
where $\sum_{y=0}^{\infty} y p_{o,r}^y = \frac{p_{o,r}}{\left(1-p_{o,r} \right)^2}$. Substituting \eqref{eq:E_X_DF2} and \eqref{eq:E_Y} into \eqref{eq:E_1alpha_DFd} and using the values of $p_{o,r}$ and $\rho$ from the proofs of Lemma \ref{L3} and Lemma \ref{L1}, respectively, $\mathbb{E}_{\mathbf{h}} \{1 - \alpha_i\}$ is given by
\begin{align}\label{eq:E_1alpha_DFd2}
      \mathbb{E}_{\mathbf{h}} \{1 - \alpha_i\} \ge \frac{2 \eta P_s e^{-\ba } }{P_r d_1^m e^{-\ba} + 2 \eta P_s}.
\end{align}

\noindent \emph{Step 4:} Finally substituting \eqref{eq:E_1alpha_DFd2} and \eqref{eq:E_g_pdt} into \eqref{eq:tau_app_d1}, we can find a lower bound on the throughput as given in \eqref{eq:tau_DFd_ana}. This completes the proof of Theorem \ref{Th4}.

\ifCLASSOPTIONpeerreview
\renewcommand{\baselinestretch}{1.0}
\fi


\end{document}